\documentclass{article}
\usepackage[utf8]{inputenc}
\usepackage{amsmath,amssymb}
\usepackage{amsfonts}
\usepackage{algpseudocode}
\usepackage{algorithm}
\usepackage{enumitem}
\usepackage[margin=1in]{geometry}
\usepackage[numbers,sort&compress]{natbib} 
\setcitestyle{numbers} 
\usepackage{authblk}

\usepackage{graphicx}
\usepackage{setspace}
\usepackage{multirow}
\usepackage{color,soul}
\usepackage{subcaption}
\usepackage{bbm}
\usepackage{booktabs}
\usepackage{siunitx}

\newcommand{\vect}[1]{\textbf{\textit{#1}}}

\newcommand{\E}{\mathrm{E}}
\newcommand{\Var}{\mathrm{Var}}

\DeclareMathOperator*{\argmax}{arg\,max}

\newtheorem{theorem}{Theorem}
\newtheorem{proposition}{Proposition}

\newlist{condenum}{enumerate}{1} 
\setlist[condenum]{label=\bfseries Condition \arabic*., ref=\arabic*, wide}

\newlist{propnum}{enumerate}{1} 
\setlist[propnum]{label=\bfseries Proposition \arabic*., ref=\arabic*, wide}

\title{Conditional variable screening for ultra-high dimensional longitudinal data with time interactions}

\author[1*]{Andrea Bratsberg}
\author[2]{Abhik Ghosh}
\author[1]{Magne Thoresen}
\affil[1]{Oslo Centre for Biostatistics and
Epidemiology, Department of Biostatistics,
University of Oslo}
\affil[2]{Indian Statistical Institute, Kolkata, India}
\affil[*]{\textit{email: a.m.bratsberg@medisin.uio.no}}

\begin{document}
\date{}
\maketitle

\begin{abstract}
In recent years we have been able to gather large amounts of genomic data at a fast rate, creating situations where the number of variables greatly exceeds the number of observations. In these situations, most models that can handle a moderately high dimension will now become computationally infeasible or unstable. Hence, there is a need for a pre-screening of variables to reduce the dimension efficiently and accurately to a more moderate scale. There has been much work to develop such screening procedures for independent outcomes. However, much less work has been done for high-dimensional longitudinal data in which the observations can no longer be assumed to be independent. In addition, it is of interest to capture possible interactions between the genomic variable and time in many of these longitudinal studies. 
In this work, we propose a novel conditional screening procedure that ranks variables according to the likelihood value at the maximum likelihood estimates in a marginal linear mixed model, where the genomic variable and its interaction with time are included in the model. This is to our knowledge the first conditional screening approach for clustered data. We prove that this approach enjoys the sure screening property, and assess the finite sample performance of the method through simulations.
\end{abstract}

\textbf{Keywords:} Interactions, linear mixed models, longitudinal analysis, sure screening property, ultra-high dimensionality, variable screening

\section{Introduction}

Emerging omics technologies have allowed us to gather an unprecedented amount of data both efficiently and in high resolution. This often gives us a situation in which the number of variables exceeds the number of observations, the so-called high-dimensional case. In order to take full advantage of the quality and depth of such data, we need suitable statistical methods. A very common assumption that is being made in order to make inferences in these situations is that the underlying structure is sparse, meaning that only a few of the variables truly have an effect on the outcome. Hence, it becomes a question of how to select these relatively few important variables out of many possible. This is the task of variable selection, and much work has been done in this area, including penalization methods such as the LASSO \citep{LASSO}, SCAD \citep{SCAD} and the Dantzig selector \citep{candes2007dantzig}, among others. These methods do variable selection in the sense that many coefficient estimates are set to exactly zero. However, when the number of variables $p$ grows non-polynomially with the number of observations $n$, we are in the ultra-high dimensional case, and these methods may become computationally infeasible or unstable. For this reason, \citet{sureindependence} proposed a two-step procedure where the first step is concerned with reducing the dimension drastically and efficiently by considering each covariate’s marginal correlation with the response, and only keeping those with high absolute marginal correlation. They called this approach Sure Independence Screening (SIS). In the second step, one can use any suitable method to perform the final variable selection. There have been many extensions and modifications to this idea. \citet{sis_in_generalized_linear_models} extended the SIS to generalized linear models by ranking the variables according to their maximum marginal likelihood estimates or the likelihood value at these estimates, the latter proposed by \citet{fan2009beyondthelinearmodel} under a general parametric framework. To model nonlinear dependencies, several different marginal utilities have been proposed, including generalized correlation \citep{hall2009generalizedcorrelation},  rank correlation \citep{li2012robust_rank_correlation} and distance correlation \citep{li2012_distancecorrelation}. \citet{fan2011NIS} proposed the nonparametric independence screening (NIS) as an extension to generalized correlation to ultra-high dimensional additive models. Within robust statistics, \citet{robustScreening} proposed a robust version of SIS and \citet{pan2018_bcorsis} used the Ball correlation to propose a robust screening procedure with minimal restrictive data assumptions. See \citet{fan2020statisticalfoundations_book} for a more comprehensive overview of the feature screening literature.

In medical studies, we are often interested in how a response changes over time, and each subject is observed at different time points. For these longitudinal data sets, observations from the same individual are likely to be more similar than observations from different individuals, which violates the assumption of independent observations made by many of the aforementioned screening procedures.  \citet{IGEE} developed a screening procedure for time course data based on generalized estimating equations (GEE) to account for such  within-subject correlation. Probably the most common way to deal with this is to introduce latent variables to model aspects of the subjects that are not captured by the observed covariates. This is the approach in mixed models, where latent random effects are included in an otherwise traditional regression model. Only recently has mixed models for high-dimensional data begun to receive attention. For example, \citet{glmmlasso} and \citet{ghoshthoresen_nonconcave} introduced respectively the LASSO and the SCAD penalty for high-dimensional linear mixed models.
However, in the ultra-high dimensional setting, these methods too become computationally infeasible. Hence, there is a need for a screening procedure in the mixed effects case for ultra-high dimensional data.

An additional issue arises when dealing with longitudinal data. We might be particularly interested in whether there are interactions between the high-dimensional covariates and time, in order to understand variation in time development. This increases the total number of potential effects even further. SIS and many of its extensions only screen for main effects. By only considering main effects in problems where we are interested in uncovering possible interactions between variables, we will likely get inaccurate results. Some selection procedures for interactions in high-dimensional data have been developed. In order to avoid the quadratic computational cost of searching among all pairs of possible interactions, most of these rely on some strong assumptions, namely the weak or strong heredity assumption, which says that for an interaction to be included, one or both of the main effects must be important (see for example \cite{bien2013hieratchichallasso, interactionscreeningforultrahighdimdata, HALXUE, ModelSelectionforHighDimensionalQuadraticRegression}). These methods are computationally feasible for high-dimensional data, but they break down if the heredity assumption is violated. Consequently, variables that only appear in pure interactions will be missed by these screening procedures.  In our setting, we are interested in a less restrictive method that allows for pure interaction effects, but on the other hand, we are only interested in interactions with time, and thus we do not have to search through all possible pairwise interactions. In principle, it would be straightforward to implement the interaction screening into SIS if we search among all possible interactions, and methods have been developed that do not need this assumption of heredity (see, for example \cite{SIRI}, \cite{kong2017interactionpursuit} and \cite{pan2018_bcorsis}). However, we are often interested in non-linear effects of time. In balanced designs, when the timing of the repeated measurements are common to all study subjects, a popular approach to analysis of non-linear effects is by treating time as a factor and introducing dummy variables. If we let the time variable be coded this way, an interaction with time will consequently consist of several parameters. This suggests the need for an extension of the SIS idea to sets of parameters.

An alternative approach to modeling  longitudinal data is through semi-parametric models. Varying-coefficient models belong to this class of models. These models may incorporate
both the within-subject correlation and interaction with time by allowing the (high-dimensional) regression coefficients to vary with time. There are various feature screening approaches based on these models (see e.g. \cite{fan2014NIS_varcoef}, \cite{song2014varying} and \cite{liu2014_varyingcoefs}). Methods for time-varying coefficient models that specifically target longitudinal data include \citet{cheng2014nonparametric}, where the NIS procedure is followed by variable selection using a refined version of the SCAD estimator for longitudinal data, \citet{chu2016feature_timevarying}
where the non-parametric approaches of \cite{fan2014NIS_varcoef} and \cite{song2014varying} are improved by incorporating within-subject dependencies, and 
\citet{zhang2019nonparametric_varying}, where the NIS idea is combined with generalized estimating equations. Other methods include \cite{niu2018nonparametric}, \cite{liu2016feature_partiallinear} and \cite{lai2020feature}.

Our focus is on the mixed model approach, and in the present work, we contribute to the current variable screening literature by proposing a screening approach based on linear mixed models that, in particular, screens for interactions with time by retaining the variables having the largest likelihood values when both main effects and interactions are included in the model. This way, we are able to screen based on the set of interaction terms by defining a marginal linear mixed model for each covariate that includes the covariate both as a main effect and as interaction with time, in addition to other possible covariates that we wish to keep out of the screening step. 
To our knowledge, no existing screening procedure is able to screen groups of variables while in addition allowing for conditioning on variables that should be kept out of the screening procedure (specifically, time in our case) in the setting of clustered data. \citet{boltssi} developed a similar likelihood-based screening procedure for two-way interactions, but they assume independent outcomes. There exist other works that address conditonal screening for general (independent) data \cite{conditional-screening,Wen_conditionalcorrelationscreening,zhou2018modelfreeconditional,Tong_conditionFDR,Cui_conditonalsurvival}. While some of these methods, e.g. \cite{Wen_conditionalcorrelationscreening}, are also defined for multivariate data,  and can in this sense be used in the setting of clustered data, they are still not fully capable of dealing with all relevant aspects of feature screeening in a longitudinal data setting.

We show that our procedure enjoys the sure screening property, meaning that, with probability tending to one, the method will capture the true active set of main effects and interactions. 
We assess the finite sample performance of the method through simulated examples, and compare the performance with the generalized estimation equations screening (GEES) approach of \cite{IGEE}, the Ball correlation screening (BCor-SIS) approach of \citet{pan2018_bcorsis} and the conditional distance correlation screening (CDC-SIS) of \citet{Wen_conditionalcorrelationscreening}. We show that for capturing groups of interaction parameters, screening on the likelihood yields the best recovery rate of the true interactions, across a wide range of settings.
Finally, we apply the proposed method to real data from a longitudinal study on measured serum triglyceride over the course of six hours, with measured mRNA on a targeted set of genes as our high-dimensional set of covariates. In this example, the ability to perform conditional screening becomes important.

\textit{Notation:} Throughout the paper, we use $(x_1,...,x_n)$ to denote a tuple of, e.g., scalars or matrices, while we use the notation $\vect{x}=[x_1,\cdots,x_n]^T$ to denote a column vector of length $n$, where $T$ (in superscript) denotes the transpose. Furthermore, the $\ell_2$ and $\ell_\infty$ norms are denoted by $\|\cdot\|$ and $\|\cdot\|_\infty$, respectively. For matrices, we denote by $\|\vect{A}\|_1$ the maximum absolute column sum of a matrix $\vect{A}$. Also, let $\boldsymbol{1}_m$ denote the $m$-vector with all entries equal to one. The rest of the notation is standard.

\section{Models and Methods}

Assume that we have observed $n$ subjects and that subject $i$ is measured at $m_i$ time points, giving a total of $N = \sum_{i=1}^{n} m_i$ observations. Denote by $\vect{y}_i$ the $m_i$-dimensional response vector of subject $i$, and by $\vect{X}_i$ the $m_i \times (\tilde{p}+1)$ design matrix of covariates, where the first column corresponds to the intercept and $\tilde{p}$ is the total number of covariates. The individuals are assumed to be independent from each other, while the measurements from the same individual are likely to be correlated. Linear mixed models allow for dependence between observations by assuming that each subject differs from each other randomly through a random effects term $\vect{Q}_i\vect{b}_i$, where $\vect{Q}_i$ is the design matrix for the random effects, and $\vect{b}_i$ is a vector of random effects, assumed to be $\mathcal{N}(0,\vect{G})$-distributed for a positive definite covariance matrix $\vect{G}$. We assume that $\vect{G}$ can be fully parameterized by a vector of variance parameters $\boldsymbol{\eta}$, so that $\vect{G} = \vect{G}(\boldsymbol{\eta})$. In general, for a linear mixed model, the response for each subject $i$ is modeled as
\begin{align}
\vect{y}_i = \textbf{X}_i\boldsymbol{\beta}+\vect{Q}_i\vect{b}_i+\boldsymbol{\epsilon}_i,
\label{EQ:linearmixedeffect1}
\end{align}
where the error term $\boldsymbol{\epsilon}_i \sim \mathcal{N}(0,\sigma_\epsilon^2\mathbb{I}_{m_i})$ and $\mathbb{I}_{m_i}$ is the $m_i \times m_i$ identity matrix and $\boldsymbol{\beta}$ is the fixed effects coefficient vector. Under this model, we know that the response vector $\vect{y}_i$ follows a Gaussian distribution, i.e. 
\begin{align*}
\vect{y}_i|\vect{X}_i, \vect{Q}_i \sim \mathcal{N}\left(\vect{X}_i\boldsymbol{\beta}, \vect{V}_i(\boldsymbol{\eta})\right),
\end{align*}
where $\vect{V}_i(\boldsymbol{\eta}) =  \vect{Q}_i\vect{G}(\boldsymbol{\eta})\vect{Q}_i^T+\sigma_\epsilon^2\mathbb{I}_{m_i}$.
Throughout the article, we will assume for simplicity that all subjects are measured $m$ times so that $m_i=m$ for all subjects $i$ and the total number of measurements becomes $N=nm$. 

\subsection{Likelihood screening}

In the setting introduced above, the observations $\{(\vect{y}_i, \vect{X}_i,\vect{Q}_i)\}_{i=1}^n$  are i.i.d realizations of the random variable 
$(\vect{y}, \vect{X}, \vect{Q})$. Moving forward, we focus on this random variable, so that we can omit the subscript $i$. We also take the conditional approach and assume that $\vect{X}$ and $\vect{Q}$ are given, and omit the explicit conditioning on these random variables for simplicity. Denote by $\boldsymbol{\beta}^*$ the true fixed effects coefficient vector.
Our main interest is the response profile as a function of time and potential interactions with time. Thus, it makes sense to screen based on a model that captures exactly this feature. In order to achieve this, we partition the design matrix for the fixed effects $\vect{X}$ into three separate design matrices; $\vect{X}_{\mathcal{M}}$ corresponding to the high-dimensional covariates, $\vect{X}_{\tau}$ corresponding to the time variable and $\vect{X}_{\mathcal{I}}$ corresponding to the interactions between the covariates and time. Similarly, we may partition (without loss of generality) the coefficient vector $\boldsymbol{\beta}^*$ into corresponding vectors $\boldsymbol{\beta}^*_{\mathcal{M}}=[\beta_{1}^*, \cdots ,\beta^*_{p}]^T,$ for the main effects,  $\boldsymbol{\beta}^*_{\mathcal{I}}=[\boldsymbol{\beta}^{*T}_{\mathcal{I}1}, \cdots,\boldsymbol{\beta}^{*T}_{\mathcal{I}p}]^T$ for interactions, where each $\boldsymbol{\beta}^*_{\mathcal{I}k}$, $k = 1,...,p$ is the vector of interaction coefficients corresponding to variable $k$, and $\boldsymbol{\tau}^*$ for the time variable. The first element of $\boldsymbol{\beta}^*$ is  $\beta_0^*$ for the intercept. 
Thus, the model \eqref{EQ:linearmixedeffect1} for a random variable $\vect{y} \in \mathbb{R}^m$ can be written as
\begin{align}
\vect{y} = \beta_0^*\vect{1}_m+\vect{X}_{\mathcal{M}}\boldsymbol{\beta}^*_\mathcal{M}+\vect{X}_{\mathcal{I}}\boldsymbol{\beta}^*_\mathcal{I}+\vect{X}_{\tau}\boldsymbol{\tau}^*+\vect{Q}\vect{b}+\boldsymbol{\epsilon},
\label{EQ:mixedmodelpartitioned}
\end{align}
where $\vect{1}_m$ is the $m$-vector of ones, and $\boldsymbol{\epsilon}\sim \mathcal{N}(0,\sigma_\epsilon^2\mathbb{I}_m)$, as in \eqref{EQ:linearmixedeffect1}.
A common crucial assumption is that the true active set of main effects and interactions, defined as $\mathcal{B} = \{1 \leq k \leq p: \lvert \beta_{k}^*\rvert\neq 0 \text{ or } \|\boldsymbol{\beta}^{*}_{\mathcal{I}k}\| \neq 0\}$, is sparse, i.e. $s = \lvert \mathcal{B} \rvert   \ll p$. We aim to estimate this set through variable screening by specifying a marginal model for each covariate $k \in \{1,...,p\}$, that includes both the main effect of the covariate $k$, the effect of time, and the interaction effect between variable $k$ and time. Let $\vect{x}_{k} \in \mathbb{R}^m$ be the random vector corresponding to the $k$th main variable, which is the $k$th column of $\vect{X}_\mathcal{M}$. The time variable can be assumed to be dummy coded with $m-1$ dummy variables corresponding to the $m$ different time points, so that $\vect{X}_{\tau} \in \mathbb{R}^{m\times m-1}$. The interaction between variable $k$ and time is then denoted by the $m\times (m-1)$ matrix $\vect{X}_{\mathcal{I}k},$ where $\vect{X}_\mathcal{I} = [\vect{X}_{\mathcal{I}1},...,\vect{X}_{\mathcal{I}p}]$.  We may note that if we have a linear effect of time, the time variable becomes a vector, say $\vect{x}_\mathcal{\tau}$, with corresponding scalar regression coefficient $\tau^*$, but we will focus on the more general case with time dummy variables in the following theoretical derivations.

\textbf{An illustrative example:}\newline
As an illustrative toy example, consider the case with $m = 4$, $p = 500$ and observations from $n = 20$ subjects. Then, for each subject $i \in \{1,...,20\}$, the observed value of the response variable $\vect{y}$ is of length $4$. For variable $\vect{x}_k$, $k \in \{1,...,500\}$, the observation for subject $i$ takes the form $\vect{x}_{ik} = [x_{i1}^k,...,x_{i4}^k]^T$ and the matrices  $\vect{X}_{i\tau}=\vect{X}_{\tau}$ (same for each subject $i$) and $\vect{X}_{i\mathcal{I}k}$ (observed value of $\vect{X}_{\mathcal{I}k}$ for subject $i$) then look like
\begin{align}
    \vect{X}_{\tau} = \begin{bmatrix}0 & 0 & 0 \\
    1 & 0 & 0\\
    0 & 1 & 0\\
    0 & 0 & 1\end{bmatrix}, \quad \text{and}\quad 
    \vect{X}_{i\mathcal{I}k} = \begin{bmatrix}0 & 0 & 0\\
    x_{i2}^k & 0 & 0 \\
    0 & x_{i3}^k & 0\\
    0 & 0 & x_{i4}^k\end{bmatrix}.
    \label{EQ:toyexample}
\end{align}
\hfill $\square$

Under the set-up above, we define the marginal maximum likelihood estimates (MLEs) for each $k \in \{1,...,p\}$ by

\begin{align}
(\widehat{\beta}_{0k}, \widehat{\beta}_{k}, \widehat{\boldsymbol{\beta}}_{\mathcal{I}k}, \widehat{\boldsymbol{\tau}}_k,\widehat{\boldsymbol{\eta}}_k) = \argmax_{(\beta_0,{\beta}_{k},\boldsymbol{\beta}_{\mathcal{I}k},\boldsymbol{\tau},\boldsymbol{\eta})} \mathbb{P}_n[l(\beta_0+\beta_{k}\vect{x}_k+\vect{X}_{\mathcal{I}k}\boldsymbol{\beta}_{\mathcal{I}k}+\vect{X}_{\tau}\boldsymbol{\tau},\boldsymbol{\eta})],
 \label{EQ:MLE}
\end{align}
where $\mathbb{P}_nf(\vect{y},\vect{X},\vect{Q}) = \frac{1}{n}\sum_{i=1}^n f(\vect{y}_i,\vect{X}_i,\vect{Q}_i)$ and 
\begin{align*}
l(\boldsymbol{\theta},\boldsymbol{\eta})=l(\boldsymbol{\theta},\boldsymbol{\eta} ; \vect{y},\vect{Q}) = -\frac{m}{2}\log(2\pi)-\frac{1}{2}\log\lvert \vect{V}(\boldsymbol{\eta})\rvert-\frac{1}{2}(\vect{y}-\boldsymbol{\theta})^T\vect{V}^{-1}(\boldsymbol{\eta})(\vect{y}-\boldsymbol{\theta}),
\end{align*}
with $\vect{V}(\boldsymbol{\eta}) =  \vect{Q}\vect{G}(\boldsymbol{\eta})\vect{Q}^T+\sigma_\epsilon^2\mathbb{I}_m$. Although we consider the log-likelihood as a function of parameters, we would like to emphasize that its value also depends on the data $(\vect{y}, \vect{X},\vect{Q})$ by explicitly mentioning it in the definition. However, to avoid confusion and for simplicity of presentation, we will omit this data dependency (in notation) in the rest of the paper, unless explicitly required.

Let us define the population parameter values corresponding to the marginal MLEs as

\begin{align}
(\bar{\beta}_{0k},\bar{\beta}_{k},\boldsymbol{\bar{\beta}}_{\mathcal{I}k},\boldsymbol{\bar{\tau}}_k,\bar{\boldsymbol{\eta}}_k) = \argmax_{(\beta_0,\beta_k,\boldsymbol{\beta}_{\mathcal{I}k},\boldsymbol{{\tau}},\boldsymbol{\eta})}\E[l(\beta_0+\beta_{k}\vect{x}_k+\vect{X}_{\mathcal{I}k}\boldsymbol{\beta}_{\mathcal{I}k}+\vect{X}_{\tau}\boldsymbol{\tau},\boldsymbol{\eta})].
\label{EQ:populationMLE}
\end{align}
We later argue that the theory holds when $\boldsymbol{\eta}$ is simultaneously estimated with the regression coefficients, and will for this reason omit the dependence on $\boldsymbol{\eta}$, and let $\vect{V}(\boldsymbol{\eta})=\vect{V}$ in the following derivations.

Since the log-likelihood function is differentiable, \eqref{EQ:populationMLE} satisfies the score equations 
\begin{align}
\E\left[\begin{bmatrix}\vect{1}_m & \vect{x}_k & \vect{X}_{\mathcal{I}k} & \vect{X}_\tau\end{bmatrix}^T\vect{V}^{-1}(\bar{\beta}_{0k}\vect{1}_m+{\bar\beta_{k}}\vect{x}_k+\vect{X}_{\mathcal{I}k}\boldsymbol{\bar{\beta}}_{\mathcal{I}k}+\vect{X}_\tau\bar{\boldsymbol{\tau}}_k)\right] \nonumber \\=
    \E\left[\begin{bmatrix}\vect{1}_m & \vect{x}_k & \vect{X}_{\mathcal{I}k} & \vect{X}_\tau \end{bmatrix}^T\vect{V}^{-1}\vect{y}\right].
    \label{scoreeqs}
\end{align}
Similarly, the MLEs in \eqref{EQ:MLE} also satisfy \eqref{scoreeqs} but with the model expectation replaced by the empirical average $\mathbb{P}_n$. The proposed likelihood screening procedure retains both the main effect and the interaction effect corresponding to the variables $k \in \{1,...,p\}$ that yield the highest log-likelihood values 
$\widehat{L}_k$, defined as

\begin{align*}
\widehat{L}_k = 
\mathbb{P}_n[l(\widehat{\beta}_{0k}+\widehat{\beta}_k\vect{x}_k+\vect{X}_{\mathcal{I}k}\widehat{\boldsymbol{\beta}}_{\mathcal{I}k}+\vect{X}_{\tau}\widehat{\boldsymbol{\tau}}_k)],
\end{align*}

where 
$(\widehat{\beta}_{0k},\widehat{\beta}_k,\widehat{\boldsymbol{\beta}}_{\mathcal{I}k},\widehat{\boldsymbol{\tau}}_k)$ is defined as in (\ref{EQ:MLE}).
We then select the variables that give the highest likelihood values when both the main effect and its interaction with time are included,
leading to the screening set
\begin{align}
    \widehat{\mathcal{B}} = \{1 \leq k \leq p: \widehat{L}_{k} \geq \widetilde{\nu}_n\},
    \label{EQ:screeningset}
\end{align}
where $\widetilde{\nu}_n$ is a predefined threshold value. This method ranks the covariates according to how much each covariate contributes to the magnitude of the likelihood function, 
similar to the likelihood ratio screening procedure in \cite{sis_in_generalized_linear_models}.  The screening set $\widehat{\mathcal{B}}$ estimates the true active set $\mathcal{B}$. In fact, Theorem 3 shows that the screening set $ \widehat{\mathcal{B}}$ is a superset of $ \mathcal{B}$, with probability tending to one.
 
\subsection{Sure screening properties}
To establish the sure screening property of the likelihood screening procedure, we make use of the ideas of the conditional screening approach of \citet{conditional-screening}. The idea is to condition on a specific set of variables which prior to screening is known to be important. Thus, the same set of variables is conditioned on in each marginal model. Since we have no interest in screening the time variable, we will condition on this variable in the marginal model corresponding to variable $k$. 
To show the sure screening property of the screening procedure \eqref{EQ:screeningset}, we must first show that the population parameters 
($\bar{\beta}_k$,$\bar{\boldsymbol{\beta}}_{\mathcal{I}k})$ are useful probes for the true marginal coefficients (${\beta}^*_{k}$,${\boldsymbol{\beta}}^{*}_{\mathcal{I}k})$. Then, we need to show that the marginal MLEs ($\widehat{\beta}_k$,$ \widehat{\boldsymbol{\beta}}_{\mathcal{I}k})$ are uniformly close to ($\bar{\beta}_k$,$\bar{\boldsymbol{\beta}}_{\mathcal{I}k})$.
We start with the sure screening properties of the population parameters.  In order to derive such theoretical properties, we define a conditional mixed linear expectation,
following the same idea from \citet{conditional-screening}, as 
	\begin{align}
    E_L(\vect{y}|\vect{H}) :=\alpha_0\vect{1} + \vect{H}\boldsymbol{\alpha},
    \label{definiton1}
\end{align}
where $(\alpha_0,\boldsymbol{\alpha})$ is the solution to 
\begin{align}
    \E\left[\begin{bmatrix}\vect{1}& \vect{H} \end{bmatrix}^T\vect{V}^{-1}(\alpha_0\vect{1}+\vect{H}\boldsymbol{\alpha})\right] = \E\left[ \begin{bmatrix}\vect{1} & \vect{H} \end{bmatrix}^T\vect{V}^{-1}\vect{y}\right],
    \label{definition2}
\end{align}
where $\vect{V} = \Var(\vect{y})$. Here, $E_L(\vect{y}|\vect{H})$ is thus the best linearly fitted regression. Now, in order to show the theoretical results, we need to define a reference level for the likelihood value. We may note that the likelihood screening approach is equivalent to a likelihood ratio screening, where we select those variables that give the largest increase in the likelihood value compared to the intercept and time effect model,
i.e., $\widehat{\mathcal{B}} = \{1 \leq k \leq p: \widehat{L}_{k}^R \geq \nu_n\}$, where $\widehat{L}_{k}^R$ is defined as
\begin{align*}
    \widehat{L}_k^R = \mathbb{P}_n\left[l(\widehat{\beta}_{0k}+\widehat{\beta}_k\vect{x}_k+\vect{X}_{\mathcal{I}k}\widehat{\boldsymbol{\beta}}_{\mathcal{I}k}+\vect{X}_{\tau}\widehat{\boldsymbol{\tau}}_k)-l(\widehat{\beta}_0^M+ \vect{X}_\tau\widehat{\boldsymbol{{\tau}}}^M)\right],
\end{align*}
and $(\widehat{\beta}_{0}^M, \widehat{\boldsymbol{\tau}}^M) = \argmax_{(\beta_0,\boldsymbol{\tau})}  \mathbb{P}_n[l(\beta_0+\vect{X}_\tau\boldsymbol{\tau})],
$ so that $\nu_n = \widetilde{\nu}_n-\mathbb{P}_n[l(\widehat{\beta}_{0}^M+\vect{X}_\tau\widehat{\boldsymbol{\tau}}^M)]$.
Clearly, the population version of $\widehat{L}_k^R$ is
\begin{align*}
     L_k^R = \E[l(\bar{\beta}_{0k}+\bar{\beta}_k\vect{x}_k+\vect{X}_{\mathcal{I}k}\bar{\boldsymbol{\beta}}_{\mathcal{I}k}+\vect{X}_{\tau}\bar{\boldsymbol{\tau}}_k)-l(\bar{\beta}_0^M+\vect{X}_\tau\bar{\boldsymbol{\tau}}^M)],
\end{align*}
where $(\bar{\beta}_0^M, \bar{\boldsymbol{\tau}}^M) = 	\argmax_{({\beta}_0,\boldsymbol{\tau})}  \E[l({\beta}_0+\vect{X}_\tau\boldsymbol{\tau})].$
Note that while we focus on interactions with time only,  $\vect{X}_\tau$ may consist of several other variables that we wish to keep out of the screening step, but we assume that the number of columns in $\vect{X}_\tau$ is small compared to the total number of covariates. Thus, for a more general notation, we will now define the matrix $\vect{Z}_k = [\vect{x}_k, \vect{X}_{\mathcal{I}k}]$, with corresponding coefficient vector $\boldsymbol{\zeta}_{k}=[{\beta}_{k},\boldsymbol{\beta}_{\mathcal{I}k}^T]^T$, and
$\vect{Z}_C = \vect{X}_\tau$ for the covariates that we condition on. We first present the relationship between the population version of $\widehat{L}_k^R$ and the conditional mixed linear expectation in the following proposition. For brevity in presentation, its proof is given in Appendix B.1; the required regularity conditions A.1 -- A.10 are presented in Appendix \ref{APP:conditions}.

\begin{proposition}
Assume that the solution to \eqref{scoreeqs} is unique and Condition A.2 and A.8 hold. Then for $k \in \{1,...,p\}$,
\begin{center}
  $L_k^R=0$ if and only if 
  $E_L(\vect{y}|\vect{Z}_C)=E_L(\vect{y}|\vect{Z}_k,\vect{Z}_C)$ 
 \end{center}
\end{proposition}

A necessary condition to ensure sure screening property at the population level is that the minimum marginal signal strength is larger than the estimation error. In the following theorem, we show that this is possible under some conditions. Its proof is in Appendix B.2. 

\begin{theorem}
Suppose that Condition A.2 and A.8 holds and that there exist constants $c_0$ and $M>0$ such that 
\begin{align}
\max_{1\leq k \leq p}\|\E(\vect{Z}_k^T\vect{V}^{-1}\vect{Z}_k)\|_1 \leq M,
	~~~~\mbox{ and } ~~~
	\min_{k\in  \mathcal{B}}\| \E[\vect{Z}_k^T\vect{V}^{-1}\{E_L(\vect{y}|\vect{Z}_k,\vect{Z}_C)-E_L(\vect{y}|\vect{Z}_C)\}\|_1
	\geq \frac{c_0\sqrt{m}}{n^{\kappa}},
\label{EQ:Cond1}
\end{align}

where $\kappa$ is as in Condition A.10. Then there exists a positive constant $c_1 > 0$ such that
\begin{align*}
    \min_{k\in\mathcal{B}}\lvert L_k^R \rvert \geq c_1mn^{-2\kappa}.
\end{align*}
\end{theorem}

In the following theorem, the sure screening property of the proposed method is stated. While we assume that the covariance structure is given, we do not need to assume that $\boldsymbol{\eta}$ is known. In fact, by utilizing Theorem 1 of \cite{bratsberg2023exponential}, we are able to show that the sure screening property of the likelihood screening will hold when $\boldsymbol{\eta}$ is estimated simultaneously with the regression coefficients. This also implies that while the estimated covariance matrix $\vect{V}$ will depend on $k$, the result will still hold. The proof of the following theorem is given in Appendix B.3.

\begin{theorem}

Assume that Conditions A.1--A.10 and the conditions in Theorem 2 hold for any $k = 1,...,p$, and assume that $n^{1-2\kappa}k_n^{-2}K_n^{-2} \to \infty$.  Then, by taking $\nu_n = c_3mn^{-2\kappa}$ for a sufficiently small positive constant $c_3$, 
there exists a constant $c_4 >0$ such that 
\begin{align*}
    \text{Pr}(\mathcal{B} \subset \widehat{\mathcal{B}}) \geq 1- s\exp(c_4n^{1-2\kappa}(k_nK_n)^{-2})+snr_1\exp(-r_0K_n^{\alpha})+r_2\exp(-r_3n),
\end{align*}
where $s = \lvert \mathcal{B} \rvert$.
\end{theorem}


\section{Numerical results}

\subsection{Threshold value}
A common challenge for all screening procedures is to select the threshold value $\tilde{\nu}_n$ such that the dimension of the fixed design matrix is reduced from the potentially very large number $p$ to a more moderate scale $d$. The threshold can be chosen in different ways, for example by controlling the false positives etc. However, in practice it is common to retain a fixed number of predictors, for example $n/\text{log}(n)$ or $(n-1)$. Our preferred approach is to use such a hard cutoff rule, followed by a regularized regression procedure where false positives can be controlled by, for example, stability selection.
In the setting of longitudinal data, it is more interesting to look at the effective sample size, which is smaller than the total number of observations since the observations from the same subject are correlated. Hence, in the following simulation examples, we set $d$ equal to the effective sample size ${n_e}$, given by

\begin{align}
    n_e=\frac{m}{1+\text{ICC}\cdot(m-1)}\cdot n,
    \label{EQ:effectivesamplesize}
\end{align}
where ICC is the intraclass correlation coefficient. In our simulations, the ICC is estimated by a simple variance component model.

\subsection{Simulation studies}

We conducted a number of simulation studies to compare the finite sample performance of the likelihood screening procedure with existing methods, specifically the GEES \cite{IGEE}, BCor-SIS \cite{pan2018_bcorsis} and CDC-SIS \cite{Wen_conditionalcorrelationscreening}. The GEES approach is developed for time course data and is therefore a relevant method for comparison. In the setting of a linear model with mean zero, the idea of \cite{IGEE} is to define the screening statistics as the $p-$dimensional vector 
\begin{align}
    \widehat{\vect{g}} = n^{-1}\sum_{i=1}^{n}\vect{X}_i^T\vect{A}_i^{1/2}\widehat{\vect{R}}^{-1}\vect{A}_i^{-1/2}\vect{y}_i,
    \label{GEESeq}
\end{align}
where $\vect{A}_i$ is an $m_i \times m_i$ diagonal matrix with the conditional variance of $\vect{y}_i$ given $\vect{X}_i$ along the diagonal, and $\widehat{\vect{R}}_i=\vect{R}_i(\hat{\rho})$ is an $m_i\times m_i$ estimated working correlation matrix that depends on a correlation parameter $\hat{\rho}$. The working correlation matrix $\vect{R}_i$ needs to be provided, and the estimate $\hat{\rho}$ is obtained via the residual-based moment method. For simplicity, we assume that the working correlation structure is the same for all subjects, so that we may omit the subscript $i$. Common working correlation structures are compound symmetry and first-order autoregressive correlation. This approach then retains the variables with largest values of $\lvert \hat{g}_k \rvert $, where $\hat{g}_k$ is the $k$th entry of  $\widehat{\vect{g}}$.
We may note that for $\vect{A}_i^{1/2}\hat{\vect{R}}^{-1}\vect{A}_i^{-1/2} = \mathbb{I}_{m}$, the procedure is equivalent to SIS of \citet{sureindependence} for linear models. 

The other two methods (BCor-SIS and CDC-SIS) are both generic screening methods in the sense that they are model-free. Both of them have some desirable properties that make them relevant for comparison with our likelihood screening procedure. BCor-SIS is a nonparametric screening procedure suitable for linear interaction models, and is in this sense relevant for comparison, although it is not developed for correlated outcomes. It is based on the Ball correlation by \citet{pan2019ballcovariance} and has been extended to be able to screen for interaction variables by considering a squared tranformation of both predictors and response within the Ball correlation measure. CDC-SIS is a model-free approach based on the conditional distance correlation by \citet{Wang_CDC}. It is able to condition on possible confounding variables and, in addition, it is defined for multivariate reponses, which makes it suitable in our case where we consider $n$ observations $\vect{y}_1,...,\vect{y}_n \in \mathbb{R}^m$.

The implementation of the GEES procedure was provided to us by the authors \cite{IGEE}. The CDC-SIS and BCor-SIS  methods are implemented using the R packages \texttt{cdcsis} and \texttt{Ball} (with \texttt{method = "interaction"} in \texttt{bcorsis}), respectively. For our method, the R package \texttt{lmer} is used to fit each marginal model.

\subsubsection*{Example 1}
We simulate a random intercept model where the high-dimensional covariates are only measured at baseline, and with linear effect of time and interactions between some of the covariates and time. That is, we simulate $n$ independent samples $\vect{y}_1,...,\vect{y}_n$ of the same length $m$ from the model \eqref{EQ:mixedmodelpartitioned} where the time variable is the vector $\vect{x}_{\tau}=[0, 1,...,m-1]^T$ and $\vect{b}_i = b_i\vect{1}_{m}$, where $b_i \sim \mathcal{N}(0,\sigma_b^2)$ and $\boldsymbol{\epsilon}_i \sim \mathcal{N}(0,\sigma_\epsilon^2\mathbb{I}_m)$. For the covariates, $X_{i1},...,X_{ip} \sim \mathcal{N}(0,0.4^2)$ and $\vect{X}_{i\mathcal{M}}$ consists of the baseline measurements repeated $m$ times. The design matrix $\vect{X}_{i\mathcal{I}}$ consists of the interactions between $\vect{X}_{i\mathcal{M}}$ and $\vect{x}_\tau$. Since we have a random intercept model, the design matrix for the random effects $\vect{Q}_i = \vect{1}_m$. We set $p =1000$, $m = 4$ and we do 400 simulations for each method. Let the true active set of variables be $\mathcal{M}_s =\{1,2,3,4\}$ for main effects and $\mathcal{I}_s = \{2,3,4,5\}$ for interaction variables, i.e. we have one pure main effect and one pure interaction effect, and the rest are both main and interaction effects.
We set the screening threshold such that the top $d=n_e$ variables are retained, where $n_e$ is calculated as in \eqref{EQ:effectivesamplesize}. We let $\beta^*_k = 1$ for $k \in \mathcal{M}_s$ and $\beta^*_{\mathcal{I}k} = 0.5$ for $k \in \mathcal{I}_s$, while the rest is zero. The intercept and time effect is set to  $\beta_0^* =0$ and $\tau^* = 0.2$, respectively.  We set $\sigma_\epsilon=0.1$
and let $\sigma_b \in \{0.1,0.9\}$ and the number of subjects $n \in \{40,80, 100\}$ to give different challenging scenarios. We compare our method to BCor-SIS, CDC-SIS and GEES with three correlation structures; independence, compound symmetry and AR(1), and refer to these three versions of GEES as SIS, GEES.cs and GEES.ar1, respectively. As GEES and CDC-SIS are not formulated for interaction screening, we implement the methods using the design matrix consisting only of the $p$ main effects. BCor-SIS is designed for detecting variables with possible linear interactions.
It is worth noting that the multivariate response formulation of CDC-SIS is currently only developed for conditioning on variables that are constant within subjects, e.g. sex or education etc. Thus, it is not directly applicable for conditioning on time, which varies within subject. For the likelihood screening procedure, we consider two dependency structures; random intercept only, and random intercept together with a random effect of time (a random slope). We call these settings LS.intercept and LS.slope, respectively. The covariance parameters $\boldsymbol{\eta}=[\sigma_b^2,\sigma_\epsilon^2]^T$ are estimated by maximum likelihood in each marginal model. Let $r_{\mathcal{M}}$ be the fraction of simulations that identifies all of  $\mathcal{M}_s$, $r_{\mathcal{I}}$ the fraction of simulations that identifies all of  $\mathcal{I}_s$, $\bar{R}_{\mathcal{M}}$ the mean recovery rate of $\mathcal{M}_s$, $\bar{R}_{\mathcal{I}}$ the mean recovery rate of $\mathcal{I}_s$ and MMS the minimum model size required to capture all of $\mathcal{M}_s$ and $\mathcal{I}_s$. Table 1 shows the results of these simulations.
First, we may note that SIS and GEES.cs gives identical results. This is because when we are only considering the baseline measurements of each covariate, the screening statistic $\widehat{g}$ with compound symmetry correlation matrix is simply a scaled version of the screening statistic corresponding to the independence structure.  Next, we note that when it comes to interactions, the likelihood screening procedures performs better than the other methods across all settings. Finally, the likelihood screening with random intercept and time slope performs significantly better than all the other methods when it comes to the main effects, while it performs similarly to LS.intercept with regards to interactions. When looking at the MMS, LS.slope performs significantly better than all other methods across all settings.

\begin{table}[H]
\centering
\caption{Table of $r_{\mathcal{M}}$,   $r_{\mathcal{I}}$, $\bar{R}_{\mathcal{M}}$ and $\bar{R}_{\mathcal{I}}$ for 400 simulations, together with the $50\%, 75\%,$ and $95\%$ percentiles of the minimum model size in Example 1.}
\begin{tabular}{lllllllll}\toprule
$n$ & Method & $r_{\mathcal{M}}$  & $\bar{R}_{\mathcal{M}}$ & $r_{\mathcal{I}}$  & $\bar{R}_{\mathcal{I}}$ & 50\% & 75\% & 95\% \\ \midrule
\multicolumn{9}{c}{$\sigma_b=0.1$} \\ \midrule
40& SIS & 0.378 & 0.820 & 0.232 & 0.779 & 276 & 587 & 916 \\
& GEES.cs & 0.378 & 0.820 & 0.232 & 0.779 & 276 & 587 & 916 \\
& GEES.ar1 & 0.375 & 0.819 & 0.235 & 0.779  & 270 & 591 & 919 \\
& BCor-SIS & 0.088 & 0.604& 0.030 & 0.554 & 576 & 800 & 965 \\
& CDC-SIS & 0.308 & 0.794 & 0.789 & 0.818 & 217 & 418 & 719 \\
 &LS.intercept & 0.082 & 0.728 & 0.670 & 0.912  & 248 & 359 & 485 \\
 & LS.slope & 0.672 & 0.914 & 0.672 & 0.914 &  18 &  66 & 278 \\
80 & SIS & 0.863 & 0.966 & 0.580 & 0.895  &  90 & 218 & 728 \\
& GEES.cs & 0.863 & 0.966 & 0.580 & 0.895 &  90 & 218 & 728 \\
& GEES.ar1 & 0.8630 & 0.9660 & 0.583 & 0.896  &  90 & 216 & 728 \\
& BCor-SIS & 0.583 & 0.891 & 0.230 & 0.796 & 344 & 587 & 880 \\
& CDC-SIS & 0.860 & 0.965 & 0.823 & 0.956 & 44 & 116 & 309 \\
& LS.intercept & 0.370 & 0.843 & 0.998 & 0.999  & 125 & 187 & 269 \\
& LS.slope  & 0.998 & 0.999 & 0.998 & 1 &  5 &  5 & 16 \\
100& SIS & 0.930 & 0.983 & 0.757 & 0.939 &  48 & 152 & 585 \\
& GEES.cs & 0.930 & 0.983 & 0.757 & 0.939 &  48 & 152 & 585 \\
& GEES.ar1 & 0.932 & 0.983 & 0.760 & 0.940 &  48 & 153 & 588 \\
& BCor-SIS & 0.750 & 0.936 & 0.328 & 0.829  & 282 & 518 & 846 \\
& CDC-SIS & 0.960 & 0.990 & 0.910 & 0.978  & 22 & 61 & 196 \\
& LS.intercept & 0.613&0.903 & 1 & 1  &94&	144&	219 \\
& LS.slope &1& 1 &1 &1 &	5& 5& 6\\
\midrule
 \multicolumn{9}{c}{$\sigma_b=0.9$} \\ \midrule
40 & SIS & 0.155& 0.679&	0.098&	0.651 	&405&688&930 \\
& GEES.cs & 0.155& 0.679&	0.098&	0.651 	&405&688&930 \\
& GEES.ar1 & 0.152 & 0.679 & 0.098 & 0.649  & 408 & 684 & 923 \\
& BCor-SIS & 0.010 & 0.404 & 0.010 & 0.377 & 662 & 839 & 975 \\
& CDC-SIS & 0.112 & 0.653 & 0.125 & 0.657  & 382 & 623 & 856 \\
& LS.intercept & 0.078 & 0.705 & 0.620 & 0.896  & 289 & 437 & 707\\
& LS.slope & 0.330 & 0.774 & 0.547 & 0.869  & 88 & 220 & 534\\
80& SIS & 0.705 & 0.925 & 0.420 & 0.853 & 184 & 441 & 850 \\
& GEES.cs  & 0.705 & 0.925 & 0.420 & 0.853 & 184 & 441 & 850 \\
& GEES.ar1 & 0.705 & 0.925 & 0.42 & 0.853  & 186 & 444 & 846 \\
& BCor-SIS & 0.280 & 0.754 & 0.110 & 0.699  & 468 & 696 & 948\\
& CDC-SIS & 0.627 & 0.905 & 0.655 & 0.910  & 111 & 241 & 608 \\
& LS.intercept & 0.260 & 0.814 & 0.995 & 0.999  & 167 & 264 & 450 \\
& LS.slope &0.920 & 0.978 & 0.985 & 0.996  & 8 & 18 & 129\\
100& SIS &0.843 & 0.961 & 0.573 & 0.891 & 120 & 340 & 859\\
& GEES.cs&0.843 & 0.961 & 0.573 & 0.891  & 120 & 340 & 859 \\
& GEES.ar1 &0.843 & 0.961 & 0.575 & 0.892  & 120 & 337 & 858 \\
& BCor-SIS &0.448 & 0.846 & 0.228 & 0.777 &  428 & 646 & 923 \\
& CDC-SIS & 0.775 & 0.944 & 0.83 & 0.958 &  63 & 164 & 462 \\
&LS.intercept &0.432 & 0.858 & 1 & 1  & 136 & 220 & 376 \\
& LS.slope & 0.975 & 0.994 & 1 & 1  & 5 & 9	 & 52\\
\bottomrule
\end{tabular}
\end{table}

\subsubsection*{Example 2}

In Example 1, we only considered the baseline measurement of the covariates. When a covariate is time-invariant like this, its effect becomes a purely between-subject effect. In order to assess performance in situations with a time-varying covariate, we now simulate $X_{i1},...,X_{ip}$ independently from a multivariate normal distribution with mean 0 and AR(1) covariance matrix with marginal variance $0.4^2$ and autocorrelation coefficient $0.8$. In addition, we let the slopes of the time variable vary randomly, so that $\vect{b}_i \sim \mathcal{N}(\textbf{0},\sigma_b^2\mathbb{I}_2)$ and $\vect{Q}_i = [\vect{1}_m, \vect{x}_\tau]$. The rest of the setup is as in Example 1. Table 2 shows the results of these simulations.
Similarly to Example 1, the two likelihood screening procedures performs best when it comes to capturing the interactions across all settings, and LS.slope also performs best when it comes to capturing the main effects. We observe an impressive effect of including a random slope in the screening, in particular for high {$\sigma_b$}. In this example, we see the benefits of incorporating the within-subject correlation, as the GEES.cs and GEES.ar1 methods performs significantly better than SIS, when focusing on both main effects and interactions. Interestingly, CDC-SIS outperforms BCor-SIS when it comes to capturing interaction effects, both in this and the previous example.

\begin{table}[H]
\centering
\caption{Table of $r_{\mathcal{M}}$,   $r_{\mathcal{I}}$, $\bar{R}_{\mathcal{M}}$ and $\bar{R}_{\mathcal{I}}$ for 400 simulations, together with the $50\%, 75\%,$ and $95\%$ percentiles of the minimum model size in Example 2.}
\begin{tabular}{lllllllll}\toprule
$n$ & Method & $r_{\mathcal{M}}$  & $\bar{R}_{\mathcal{M}}$ & $r_{\mathcal{I}}$  & $\bar{R}_{\mathcal{I}}$ & 50\% & 75\% & 95\% \\ \midrule
\multicolumn{9}{c}{$\sigma_b=0.1$} \\ \midrule
40& SIS & 0.665 & 0.914 & 0.450 & 0.861  & 166 & 352 & 794 \\
& GEES.cs &	0.715 & 0.928 & 0.562 & 0.887 & 89 & 257 & 727\\
& GEES.ar1 & 0.835 & 0.959 & 0.632 & 0.908 & 51 & 169 & 619\\
& BCor-SIS & 0.448 & 0.841 & 0.168 & 0.760  & 370 & 648 & 900 \\
& CDC-SIS  & 0.370 & 0.820 & 0.310 & 0.808 & 224 & 470 & 799 \\
& LS.intercept & 0.647 & 0.911 & 0.762 & 0.941 & 59 & 175 & 489\\
& LS.slope & 0.945 & 0.986 & 0.820 & 0.955 & 14 & 51 & 213\\
80& SIS &0.958 & 0.989 & 0.828 & 0.957  & 29 & 103 & 548\\
& GEES.cs & 0.988 & 0.997 & 0.897 & 0.974 & 13 & 49 & 227\\
& GEES.ar1 & 0.993 & 0.998 & 0.932 & 0.983  & 8 & 25 & 176\\
& BCor-SIS & 0.915 & 0.979 & 0.470 & 0.868 & 149 & 418 & 786\\
& CDC-SIS & 0.895 & 0.974 & 0.762 & 0.941 & 54 & 167 & 503 \\
& LS.intercept & 0.980 & 0.995 & 0.985 & 0.996 & 9 & 20 & 96\\
& LS.slope & 1 & 1 & 0.995 & 1 &  5 &  6 & 25 \\
100 & SIS &	0.993 & 0.998 & 0.917 & 0.979 & 13 & 50 & 273\\
& GEES.cs & 0.998 & 0.999 & 0.965 & 0.991 & 7 & 17 & 114\\
& GEES.ar1 & 1 & 1 & 0.975 & 0.994  & 6 & 12 & 82\\
& BCor-SIS& 0.960 & 0.990 & 0.623 & 0.906  & 99 & 298 & 767 \\
& CDC-SIS& 0.958 & 0.989 & 0.865 & 0.966  & 30 & 94 & 337 \\
& LS.intercept& 1 & 1 & 0.998 & 0.999  &  6 &  9 & 30 \\
& LS.slope& 1 & 1 & 1 & 1  & 5 & 5 & 7 \\
\midrule
\multicolumn{9}{c}{$\sigma_b=0.9$} \\ \midrule
40& SIS & 0.105 & 0.616 & 0.058 & 0.587 & 488 & 734 & 938 \\
& GEES.cs & 0.117 & 0.628 & 0.098 & 0.605 & 503 & 718 & 952\\
& GEES.ar1 & 0.222 & 0.724 & 0.132 & 0.686 & 400 & 661 & 941\\
& BCor-SIS & 0.030 & 0.454 & 0.005 & 0.424 & 679 & 826 & 953\\
& CDC-SIS & 0.022 & 0.415 & 0.018 & 0.401 & 608 & 790 & 948 \\
& LS.intercept & 0.060 & 0.574 & 0.108 & 0.614 & 440 & 680 & 922\\
& LS.slope & 0.815 & 0.954 & 0.662 & 0.914 & 50 & 160 & 510\\
80 & SIS & 0.570 & 0.884 & 0.382 & 0.834& 282 & 568 & 915\\
& GEES.cs & 0.637 & 0.904 & 0.450 & 0.852 & 213 & 510 & 804\\
& GEES.ar1 & 0.780 & 0.943 & 0.540 & 0.882  & 157 & 370 & 816\\
& BCor-SIS & 0.338 & 0.776 & 0.160 & 0.701  & 446 & 663 & 922\\
& CDC-SIS  & 0.272 & 0.756 & 0.208 & 0.736  & 402 & 594 & 873 \\
& LS.intercept  & 0.535 & 0.876 & 0.578 & 0.887 & 193 & 394 & 813\\
& LS.slope & 0.995 & 0.999 & 0.950 & 0.988  & 6 & 13 & 131\\
100 & SIS & 0.790 & 0.946 & 0.510 & 0.873  & 208 & 467 & 898\\
& GEES.cs & 0.838 & 0.958 & 0.593 & 0.894  & 134 & 362 & 856\\
& GEES.ar1 & 0.890 & 0.973 & 0.650 & 0.911  & 93 & 262 & 780\\
& BCor-SIS & 0.552 & 0.873 & 0.280 & 	0.783 & 437 & 690 & 939\\
& CDC-SIS & 0.525 & 0.864 & 0.385 & 0.820 & 316 & 530 & 866 \\
&LS.intercept & 0.785 & 0.946 & 0.757 & 0.938 &  114 & 288 & 670\\
& LS.slope & 1 & 1 & 0.985 & 0.996 &  5 &  7 & 34 \\
\bottomrule
\end{tabular}
\end{table}

\subsubsection*{Example 3}
In order to assess the performance of the methods in a situation with non-linear effects of time, we simulate from a random intercept model as in Example 1, but now the time variable is a dummy variable, i.e., $\vect{X}_\tau$ and $\vect{X}_{i\mathcal{I}k}$ are as in \eqref{EQ:toyexample} for each subject $i$, while keeping $\vect{Q}_i$, $\vect{b}_i$, $\sigma_\epsilon$, $\mathcal{M}$ and $\mathcal{I}$ as in Example 1. We set $\beta_0^* =0$, $\boldsymbol{\tau}^*=  \vect{1}_4$, $\boldsymbol{\beta}_{\mathcal{I}k}^* = \vect{1}_4$ for $k \in \mathcal{I}_s$, and $\beta^*_k = 1$ for $k \in \mathcal{M}_s$, and zero otherwise. We consider only baseline measurements of the high-dimensional covariates as in Example 1. As seen in Example 1, GEES with compound symmetry correlation structure is equivalent to that of an independence structure. Thus, we omit the results for GEES.cs here. The results are given in Table \ref{Fig:example3}. Again focusing on the interactions, we observe that the likelihood screening approach outperforms the other methods across all settings, suggesting that screening on the likelihood value is a good approach in cases where we have a set of interaction parameters. Similarly to Example 1 and 2, we see the benefits of incorporating a random slope in the likelihood screening for capturing the main effects.

\begin{table}[H]
\centering
\caption{Table of $r_{\mathcal{M}}$,   $r_{\mathcal{I}}$, $\bar{R}_{\mathcal{M}}$ and $\bar{R}_{\mathcal{I}}$ for 400 simulations, together with the $50\%, 75\%,$ and $95\%$ percentiles of the minimum model size in Example 3.}
\begin{tabular}{lllllllll}\toprule
$n$ & Method & $r_{\mathcal{M}}$  & $\bar{R}_{\mathcal{M}}$ & $r_{\mathcal{I}}$  & $\bar{R}_{\mathcal{I}}$ & 50\% & 75\% & 95\% \\ \midrule
\multicolumn{9}{c}{$\sigma_b=0.1$} \\ \midrule
40 & SIS& 0.382 & 0.821 & 0.238 & 0.781  & 276 & 587 & 916 \\
& GEES.ar1 & 0.420 & 0.836 & 0.198 & 0.770  & 306 & 607 & 909\\
& BCor-SIS & 0.092 & 0.614 & 0.043 & 0.583  & 558 & 805 & 958 \\
& CDC-SIS & 0.312 & 0.797 & 0.235 & 0.775 & 266 & 505 & 775 \\
& LS.intercept & 0.090 & 0.730 & 0.675 & 0.914 & 251 & 361 & 480 \\
& LS.slope &0.495 & 0.858 & 0.767 & 0.940 & 49 & 84 & 261\\
80 & SIS & 0.865 & 0.966 & 0.588 & 0.897  &  90 & 218 & 728 \\
& GEES.ar1 & 0.905 & 0.976 & 0.507 & 0.877  & 112 & 276 & 835 \\
& BCor-SIS & 0.555 & 0.884 & 0.310 & 0.818  & 294 & 561 & 873 \\
& CDC-SIS & 0.843 & 0.961 & 0.728 & 0.932  & 61 & 156 & 424 \\
& LS.intercept & 0.380 & 0.845 & 0.998 & 0.999 & 125 & 186 & 271\\
& LS.slope & 0.995 & 0.999 & 0.995 & 0.999  & 7 & 11 & 28 \\
100 & SIS & 0.935 & 0.984 & 0.762 & 0.941  & 48 & 152 & 585 \\
& GEES.ar1 &0.955 & 0.989 & 0.660 & 0.915 &  66 & 186 & 659\\
& BCor-SIS &0.720 & 0.929 & 0.422 & 0.853  & 234 & 494 & 845 \\
& CDC-SIS & 0.945 & 0.986 & 0.863 & 0.966  & 33 & 89 & 304 \\
& LS.intercept & 0.632 & 0.908 & 1 & 1 &   92 & 144 & 219 \\
& LS.slope & 1 & 1 & 1 & 1 &  6 &  7 & 13 \\\midrule
\multicolumn{9}{c}{$\sigma_b = 0.9$} \\ \midrule
40& SIS & 0.155 & 0.682 & 0.098 & 0.654  & 405 & 688 & 930\\
& GEES.ar1 & 0.155 & 0.676 & 0.080 & 0.634 & 432 & 690 & 952\\
& BCor-SIS &0.018 & 0.431 & 0.013 & 0.410  & 650 & 831 & 964\\
& CDC-SIS & 0.110 & 0.646 & 0.102 & 0.638  & 434 & 661 & 878 \\
& LS.intercept & 0.068 & 0.706 & 0.623 & 0.897 & 286 & 436 & 723 \\
& LS.slope & 0.178 & 0.737 & 0.672 & 0.911  & 144 & 320 & 665\\
80  & SIS &0.708 & 0.926 & 0.422 & 0.853  & 184 & 441 & 850\\
& GEES.ar1 &0.735 & 0.931 & 0.348 & 0.832  & 230 & 517 & 884\\
& BCor-SIS & 0.290 & 0.771 & 0.162 & 0.728  & 438 & 693 & 951 \\
& CDC-SIS & 0.627 & 0.905 & 0.575 & 0.89 & 128 & 287 & 692 \\ 
& LS.intercept & 0.272 & 0.818 & 0.993 & 0.998 & 166 & 263 & 455\\
& LS.slope & 0.787 & 0.946 & 0.993 & 0.998 & 36 & 76 & 237\\
100 & SIS & 0.843 & 0.961 & 0.578 & 0.892 & 120 & 340 & 859 \\
& GEES.ar1 & 0.863 & 0.966 & 0.460 & 0.863  & 168 & 429 & 878 \\ 
& BCor-SIS & 0.470 & 0.857 & 0.280 & 0.799 & 382 & 622 & 905 \\
& CDC-SIS & 0.765 & 0.941 & 0.757 & 0.939 & 78 & 217 & 542 \\
& LS.intercept & 0.440 & 0.860 & 1 & 1 & 132 & 216 & 383 \\
& LS.slope & 0.907 & 0.977 & 1 & 1 & 19 &  47 & 164 \\\bottomrule
\end{tabular}
\label{Fig:example3}
\end{table}

In the Supplementary material, we have performed the same simulation studies as in Example 1, 2 and 3 for both disjoint sets, i.e.  $\mathcal{M}_s = \{1,2,3,4\}$ and  $\mathcal{I}_s =  \{5,6,7\}$, and for identical sets $\mathcal{M}_s =\mathcal{I}_s = \{1,2,3,4\}$. In these results, we see similar trends, that both of the likelihood screening methods performs better than the other methods when it comes to capturing the interactions. In addition, LS.slope greatly outperforms the other methods across many settings with respect to capturing the main effects. This is particularly apparent when considering the MMS. However, LS.slope performs a bit worse with identical sets and the set-up under Example 1 (a random intercept model). This is not surprising as the LS.slope is a misspecification of the correlation structure when the simulated model has only a random intercept.

\subsubsection*{Example 4}

In the previous examples, we assumed that the columns of the design matrix $\vect{X}$ were uncorrelated. However, in real-world scenarios, variables are typically correlated, and we expect the level of correlation to affect all univariate screening methods. To briefly explore the potential impact of multicollinearity, we replicate the setup of Example 1, but now let the $p$ variables (for each subject and each time point) come from a zero-mean multivariate normal distribution whose covariance matrix $\boldsymbol{\Sigma}$ equals $1$ along the diagonal, while the off-diagonal entries equal $\omega \in \{0, 0.5, 0.9\}$. We set $n = 80$ and $p = 1000$ and report $r_\mathcal{M}$ and $r_\mathcal{I}$, i.e., the probability of including the true active sets $\mathcal{M}_s$ and $\mathcal{I}_s$, respectively.

\begin{table}[H]
\centering
\caption{$r_\mathcal{M}$ and $r_\mathcal{I}$ for different levels of between-variables correlation $\omega \in \{0, 0.5, 0.9\}$ for Example 4.}
\begin{tabular}{lllllll}\toprule Method &  \multicolumn{3}{c}{$r_\mathcal{M}$} & \multicolumn{3}{c}{$r_\mathcal{I}$}\\[5px]
\cmidrule(lr){2-4}\cmidrule(lr){5-7} & {$\omega = 0$} & {$\omega = 0.5$}&{$\omega = 0.9$} & {$\omega = 0$}& {$\omega = 0.5$}&{$\omega = 0.9$}\\
\midrule SIS& 0.833  & 0.627& 0.560 & 0.593  & 0.507& 0.372\\
GEES.ar1& 0.835 & 0.627 & 0.565 & 0.593  & 0.507 & 0.375 \\
BCor-SIS & 0.595 & 0.492 & 0.378 & 0.195 & 0.302 & 0.202\\
CDC-SIS & 0.853 &0.485& 0.390 & 0.818 & 0.485 & 0.372\\
LS.intercept & 0.400 & 0.165 &0.135 & 0.998 &  0.890 & 0.800\\
LS.slope & 0.998 &0.948 &0.912 &  0.998 &0.948 & 0.912\\
\bottomrule
\end{tabular}
\end{table}

Similar to what is reported in \citet[Section 4.2.1]{sureindependence}, we see that multicollinearity, indicated by $\omega = 0.9$, undermines the performance of all methods. Still, the LS.slope method consistently outperforms the others in capturing both main effects and interactions across different values of $\omega$.

\subsection{Real life data example}
\label{section:reallifeexample}

Elevated serum triglyceride (TG) levels are known to be associated with the risk of cardiovascular disease (CVD), and the CVD risk reducing effect of marine omega-3 fatty acids is believed to be mainly mediated by reduction of TG levels. However, it is well known that there is large individual variation with regard to TG response in relation to intake of dietary fat. In this example, we will analyze data from a randomized controlled cross-over trial in $n=43$ healthy subjects, males and females, age 25 to 46 years, with mean body mass index 23.6 \si[per-mode=symbol]{\kilogram\per\square\meter}\citep{HanssonTriglycerid}. The subjects were exposed to four different meals with similar amounts of fat from different dairy products, and the response was serum concentration of TG measured before the meal and 2, 4, and 6 hours after. The original aim of the study was to compare the effect of the four different meals, on TG response. In addition to the primary exposure (meal), we have measured mRNA on a targeted set of genes before each meal. Our primary interest is if the TG response is related to mRNA, i.e. if there are any interactions between some genes and time. As we have mRNA from a total of 624 genes, it makes sense to perform some variable screening to identify promising candidate genes. The TG response after each meal is typically highly non-linear and a practical solution to the analysis is to introduce dummy variables for time. 

We apply our suggested conditional screening method, conditioning on meal in addition to time and include a random intercept and random effect of time, as LS.slope in the simulation studies. The variance parameters are estimated by MLE. We then retain the top $n_e$ variables, where the effective sample size is now estimated by fitting a random intercept model and conditioning on meal, leading to $n_e \approx 61$. 

For comparison, we include GEES (with three working correlation structures, as before) and BCor-SIS, although these methods do not formally allow for conditional screening, which is necessary in order to take care of the meal effect. We also include a comparison with the CDC-SIS, conditioning on meal.

\begin{figure}[H]
\caption{Sorted histogram of the selected genes by the likelihood screening method in 100 bootstrap samples.}
\centering
\includegraphics[scale=0.75]{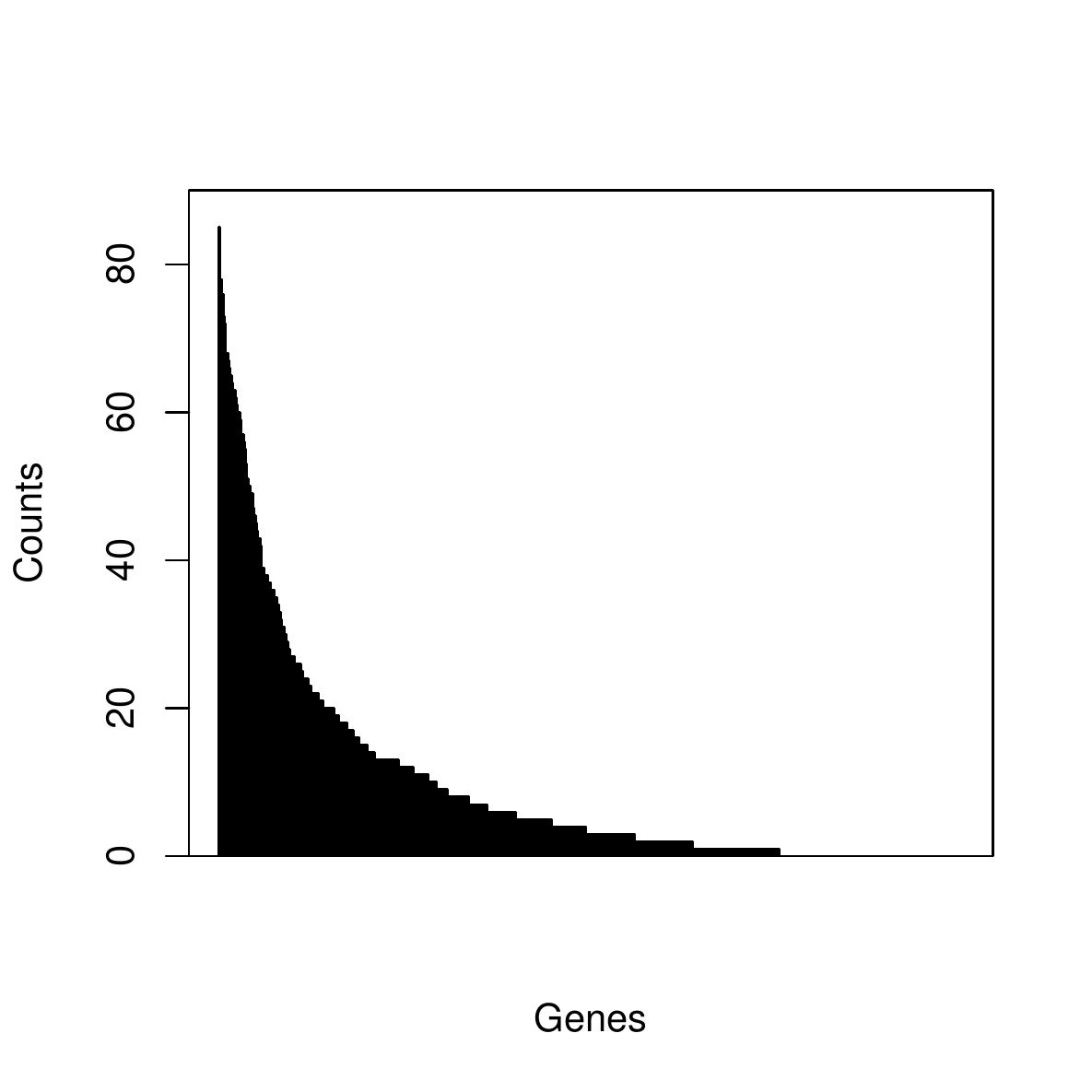}
\label{Fig:histogram}
\end{figure}

In order to investigate the variability of the different screening procedures, we also include a bootstrapping step, where we apply the screening procedures on random subsamples of size $n$ where the sampling is done with replacement. We repeat this $100$ times. Figure \ref{Fig:histogram} shows the distribution of the screening frequencies for the likelihood screening, i.e. the number of times each variable was retained among the top $n_e$ variables, in order to illustrate the variation among the estimated active sets across the samples. This pattern was quite similar across all screening methods, with an exception for SIS which showed an even larger variation. For likelihood screening, in total, $470$ variables were retained in at least one subsample, out of $624$,
indicating the need for a liberal screening threshold.

For all screening methods, we complete the analysis by performing a final variable selection with the SCAD procedure of \citet{ghoshthoresen_nonconcave}. We first do a single pre-screening and retain the top $n_e$ variables. It is interesting to investigate the overlap between the screening sets based on the different methods. Among the $n_e$ selected variables, the number of overlapping variables between the different methods is reported in Table 5. 

\begin{table}[H]
\centering
\caption{Table of the number of variables in the intersection of the screening sets based on the methods LS.slope, SIS, GEES.cs, GEES.ar1, BCor-SIS, CDC-SIS.}
\begin{tabular}{ccccccc}
\toprule
&LS.slope &SIS & GEES.cs & GEES.ar1 & BCor-SIS & CDC-SIS\\
 \midrule
LS.slope  & - & 12 
 & 29 &  6 & 12 & 18 \\
\midrule
SIS   & &- & 8 & 0 & 27 & 42 \\
\midrule
GEES.cs  &  &  &-& 8 & 5 & 10\\
\midrule
GEES.ar1 &  &  &   &-& 0 & 1\\
\midrule
BCor-SIS & & & & &-&  35\\
\midrule
CDC-SIS & & &&&& -\\
\bottomrule
\end{tabular}
\end{table}

One potential reason for the lack of overlap might be high correlation among the variables, as discussed in Section 3.2, Example 4.
Note that while the GEES.cs has 8 common variables with both SIS and GEES.ar1, these are not the same. Finally, based on the $n_e$ selected variables, we fit a SCAD model with these variables as possible main effects and interaction variables (with time) and with a regularization parameter selected through cross-validation. For the final fit, we keep time and meal unpenalized in the model, and fit a random intercept model. With the given regularization parameter, stability selection based on complementary pairs \cite{stabilityselectionShah} was performed, in which we take a random subsample of size $\lfloor n/2 \rfloor$ of the original data set and perform a variable selection with SCAD on the two disjoint subsets. We performed the resampling 50 times. 

Among the six different methods, only the likelihood screening, the GEES.ar1 and the BCor-SIS were able to capture interaction variables and Table 6 shows the main effects and interactions whose estimated coefficients were non-zero in more than $60 \%$ of the subsamples (the relative selection frequency in parentheses) for these three methods. Finally, to have a comparison between these three methods, we fit a linear mixed model based on the selected sets shown in Table 6 and report the mean prediction error from 100 random splits of the original data.

\begin{table}[H]
\centering
\caption{Table of the relative selection frequency of the top selected main variables and interaction variables based on the SCAD method for linear mixed models, and the mean prediction error (PE) from a linear mixed model with these selected variables. The mean PE is based on 100 random splits of the data.}
\begin{tabular}{llll}\toprule
Method & Main variables  & Interaction variables & PE \\ \midrule
LS.slope &  \begin{tabular}{@{}l@{}} BLNK (0.81), FCER1A (0.80), \\ ITGA2B (0.80), ATG16L1 (0.74),\\ LILRA3 (0.68), HLA.DRB1 (0.67)
 \end{tabular} & \begin{tabular}{@{}l@{}}GUSB (0.71) \end{tabular} & 0.286 \\
\midrule
GEES.ar1 &  \begin{tabular}{@{}l@{}} TRAF5 (0.85),  	sCTLA4 (0.68), \\
  ICAM2 (0.67), 	ZEB1 (0.67), \\ CD209 (0.64), 	CTLA4\_all (0.61), \\
  IL1RL2 (0.60)
 \end{tabular} & CTLA4\_all (0.69) & 0.304 \\
 \midrule
BCor-SIS &\begin{tabular}{@{}l@{}} LILRA3 (0.75),  TNFRSF4(0.74),\\  FCER1A (0.71),
  TCF4 (0.68),\\ HLA.DRB1 (0.66), KIR3DL2 (0.60)
 \end{tabular} & \begin{tabular}{@{}l@{}} HLA.DRB1 (0.62) \end{tabular} & 0.284\\
\bottomrule
\end{tabular}
\end{table}

Looking at Table 6, we notice the limited overlap between the three selected models. If we take a closer look at the likelihood screening, we see that six main effect variables were selected in more than $60\%$ of the subsamples, while one interaction was selected. The GUSB gene only appears through an interaction with time. The interpretation of this is that the gene has no relation to fasting TG level, but is related to TG response to the given meal. To what extent this is biologically plausible is beyond the scope of the current paper. In terms of the prediction errors, there is little difference between the three methods. As a general comment, the large variability in the screening sets, the limited overlap between the methods, and the similar prediction performance of the final methods indicate that there is limited evidence for a sparse set of genes playing a major role in TG response.

\section{Discussion}
The need for interaction screening procedures for response profiles as a function of time in high-dimensional longitudinal data motivated this work. We have proposed a conditional screening procedure that screens for both main effects and interactions, without the need for the heredity assumption, and that is able to capture interactions consisting of several terms. To the best of our knowledge, this is the first conditional screening procedure for clustered data. We have shown the sure screening properties of the suggested method. We compared the finite sample performance of the method with the GEES \cite{IGEE}, BCOR-SIS \cite{pan2018_bcorsis} and CDC-SIS \cite{Wen_conditionalcorrelationscreening} through simulations. While GEES is not developed to screen for interactions directly, it is suitable for correlated outcomes. On the other hand, BCor-SIS is not formulated for within-cluster correlations, but it is developed for interaction screening. Finally, CDC-SIS \cite{Wen_conditionalcorrelationscreening} is formulated for conditional screening and for multivariate outcomes. We saw that the likelihood screening approach is better for capturing the interactions across a range of different signal strengths. The likelihood screening with random slope in time in addition to random intercept also performed significantly better than the other methods when it comes to capturing main effects in many set-ups. When focusing on the MMS, LS.slope greatly outperforms the competing methods. Even though the correlation structure is wrongly specified in the setting of a random intercept data-generating process, it still performs better than the correctly specified LS.intercept. This is due to the unexplained variation induced by other active interaction terms not included in the present marginal model, which is better captured by a random slope model. 

As mentioned in Section 3.2, Example 4, the presence of multicollinearity raises some challenges for univariate screening procedures. For instance, covariates that are marginally uncorrelated but jointly correlated with the response, will likely be overlooked by screening methods that rely solely on marginal information of the covariates. For this reason, \citet{sureindependence} proposes an iterative version of the screening, in which an initial active set is estimated by variable screening with the original response, while in the subsequent iterations the residuals from the previous model is used as the new response to obtain a new estimated active set, which is then added to the previous set. This iterative approach is straightforward to implement for the proposed procedure and can be expected to have an improved performance, similar to what is seen for SIS, GEES and BCor-SIS.

How to set the screening threshold is crucial. We argue that it makes most sense to use a liberal threshold, using the screening procedure primarily to reduce the number of variables to a level that is manageable and leave the final variable selection to some regularized regression procedure. Motivated by this, we used a threshold equal to $d=n_e$. 

Finally, this work focuses on the linear mixed model. To extend it to generalized linear mixed models would be of interest, but outside the scope of this paper.

\section*{Acknowledgments}

The authors wish to thank the Editor, Associate Editor and the reviewers for their valuable comments which have led to a significant improvement of this paper. In addition, the authors would like to thank Prof. Stine Ulven, Department of Nutrition, University of Oslo, for providing the triglyceride data used in the example.

\section*{Data accessibility}
The data used in this analysis are from a third party and is not available for distribution. 

\section*{Declaration of conflicting interests}
The authors declared no potential conflicts of interest with respect to the research, authorship, and/or publication of this
article.

\section*{Funding}
The research of AG is partially supported by a research grant (No.  SRG\/2020\/000072) from the Science \& Engineering Research Board (SERB), Government of India, India.

\newpage

\bibliographystyle{unsrtnat}
\bibliography{references}

\newpage

\appendix
\section{Regularity conditions for sure screening properties}
\label{APP:conditions}

We consider the following conditions for every $k = 1,...,p$ while proving the sure screening property for the proposed screening procedure based on likelihood value.

\begin{itemize}
	\item[\textit{A.1}] Suppose that there exists a suitable constant $Z>0$ such that  	$(\bar{\beta}_{0k},\boldsymbol{\bar{\zeta}}_{k},\boldsymbol{\bar{\tau}}_k)$ is an interior point of the compact and convex set 
	$
	\boldsymbol{\mathcal{Z}} = \left\{({\beta}_0,\boldsymbol{\zeta}_{k},\boldsymbol{\tau}): \lvert{\beta}_0-\bar{\beta}_{0k}\rvert+\|\boldsymbol{\zeta}_{k}-\boldsymbol{\bar{\zeta}}_{k}\|_1+\|\boldsymbol{\tau}-\boldsymbol{\bar{\tau}_{k}}\|_1\rvert< Z\right\}.
	$
	\item[\textit{A.2}] There exists a positive constant $C_0$ such that
	$\E \|\boldsymbol{1}_m^T[\vect{Z}_{k},\vect{Z}_C]\|^2 \leq m C_0$.
	\item[\textit{A.3}] 

 The variance-covariance matrix $\vect{V}(\boldsymbol{\eta})$ is positive definite at $\boldsymbol{\eta} = \bar{\boldsymbol{\eta}}_k$ and the matrix $[\vect{1},\vect{Z}_k,\vect{Z}_C]$ is of full column rank. Additionally,   $\|I_{k}\|$ is bounded from above, where $I_{k}$ is the marginal Fisher information matrix, given by
    \begin{align*}
        I_{k}= \E\left[[\vect{1},\vect{Z}_k,\vect{Z}_C]^T\vect{V}^{-1}(\bar{\boldsymbol{\eta}}_k)[\vect{1},\vect{Z}_k,\vect{Z}_C]\right].
    \end{align*} 
    	\item[\textit{A.4}]
	There exist positive constants $s_0, s_1$ such that
\begin{align*}
	&\sum_{j=1}^m\Bigg[\E\Big\{\exp\big[(\vect{x}_j^T\boldsymbol{\beta}^*+\vect{q}_j^T\vect{b}+s_0)^2/2-(\vect{x}_j^T\boldsymbol{\beta}^*+\vect{q}_j^T\vect{b})^2/2\big]\Big\}
	\\
	&~~~~~~~~~~~~~~~+ \E\Big\{\exp\big[(\vect{x}_j^T\boldsymbol{\beta}^*+\vect{q}_j^T\vect{b}-s_0)^2/2-(\vect{x}_j^T\boldsymbol{\beta}^*+\vect{q}_j^T\vect{b})^2/2\big]\Big\}\Bigg] \leq s_1,
\end{align*}
where $\vect{x}_j^T$ and $\vect{q}_j^T$ are the $j$th row of $\vect{X}$ and $\vect{Q}$, respectively. 
	\item[\textit{A.5}]
	There exist positive constants $r_0, r_1, \alpha$ (independent of $k$) such that, for a sufficiently large $t$, we have
	\begin{align*}
	\text{Pr}(\| \mathbb{Z}_k\|_1 > t) \leq (r_1-s_1)\exp(-r_0t^\alpha),
	\end{align*}
	 where $\mathbb{Z}_k:=[\vect{1},\vect{Z}_k,\vect{Z}_C]$ and $s_1$ is the same constant from Condition A.4.
	\item[\textit{A.6}]
 For a given $({\beta}_0,\boldsymbol{\zeta}_{k},\boldsymbol{\tau})\in \boldsymbol{\mathcal{Z}}$, 
	the function $l(\cdot)$ satisfies the Lipschitz condition with a positive constant $k_n$ at $\boldsymbol{\eta} = \bar{\boldsymbol{\eta}}_{k}$.
	That is, for any $({\beta}_{0},\boldsymbol{\zeta}_{k},\boldsymbol{\tau}),({\beta}'_{0},\boldsymbol{\zeta}'_{k},\boldsymbol{\tau}')\in \boldsymbol{\mathcal{Z}}$
	and all $(\vect{y},\vect{X})\in\Lambda_n :=  \{(\vect{y},\vect{X}): 
\|\vect{X}\|_1 \leq K_n, \| \vect{y}\|_\infty\leq K_n^*\}$ for sufficiently large positive constants $K_n$ and $K_n^*$, we have 
	\begin{align}
&\left\lvert l\left({\beta}_{0}+\vect{Z}_k\boldsymbol{\zeta}_k+\vect{Z}_C\boldsymbol{\tau},\bar{\boldsymbol{\eta}}_{k}\right) - 
l\left({\beta}_{0}'+\vect{Z}_k\boldsymbol{\zeta}_k'+\vect{Z}_C\boldsymbol{\tau}',\bar{\boldsymbol{\eta}}_{k}\right)
	\right\rvert 	\\ &\leq k_n\left\lvert m(\beta_{0}-\beta_{0}')+\vect{1}_m^T\vect{Z}_{k}(\boldsymbol{\zeta}_k-\boldsymbol{\zeta}_k')+\vect{1}^T_m\vect{Z}_C(\boldsymbol{\tau}-\boldsymbol{\tau}')
	\right\rvert.
	\nonumber
	\end{align}
	\item[\textit{A.7}]
	There exists an $\epsilon_1 > 0$ (independent of $k$) such that, for the constant $K_n$ in Condition A.6, we have
    \begin{align*}
        \sup_{\substack{\{{\beta}_{0},\boldsymbol{\zeta}_{k},\boldsymbol{\tau}\}\in \boldsymbol{\mathcal{Z}},\\\|[{\beta}_0,\boldsymbol{\zeta}_{k}^T,\boldsymbol{\tau}^T]^T-[\bar{\beta}_{0k},\bar{\boldsymbol{\zeta}}_{k}^T,\boldsymbol{\bar{\tau}}_k^T]^T\| \leq \epsilon_1}}
        \left\lvert \E[(\boldsymbol{\mu}^T\mathbb{Z}_k\vect{V}^{-1}(\bar{\boldsymbol{\eta}}_k)\mathbb{Z}_k\boldsymbol{\mu})I(\|\mathbb{Z}_k\|_1> K_n)] \right\rvert \leq o(m/n),
    \end{align*}
    where $\boldsymbol{\mu}:=[{\beta}_{0},\boldsymbol{{\zeta}}_{k}^T,\boldsymbol{\tau}^T]^T$ and $\mathbb{Z}_k$ is as in Condition A.5. 
	\item[\textit{A.8}]
	There exists a positive constant $\tilde{K}$ (independent of $k$) such that, 
	for all $({\beta}_0,\boldsymbol{\zeta}_{k},\boldsymbol{\tau})\in \boldsymbol{\mathcal{Z}}$, we have 
    \begin{align*}
    \E\left[l(\bar{\beta}_{0k}+\vect{Z}_k\boldsymbol{\bar{\zeta}}_{k}+\vect{Z}_C\boldsymbol{\bar{\tau}}_{k},\bar{\boldsymbol{\eta}}_{k})-l({\beta}_{0}+\vect{Z}_k\boldsymbol{\zeta}_k+\vect{Z}_C\boldsymbol{\tau},\bar{\boldsymbol{\eta}}_{k})\right] 
        \\\geq \tilde{K} \left\|[{\beta}_{0},\boldsymbol{\zeta}_k^T,\boldsymbol{\tau}^T]^T-[\bar{\beta}_{0k},\boldsymbol{\bar{\zeta}}_{k}^T,\boldsymbol{\bar{\tau}}_{k}^T]^T\right\|^2.
    \end{align*}
\item[\textit{A.9}]
Suppose that there exist constants $r_2$ and $r_3$ such that
$\text{Pr}(\Omega_n^c) \leq r_2\exp(-r_3n)$, where
$\Omega_n = \{(\vect{y},\vect{X}):  \mathbbm{P}_nl(\widehat{\beta}_{0k}+\vect{Z}_k\widehat{\boldsymbol{\zeta}}_k+\vect{Z}_C\widehat{\boldsymbol{\tau}}_{k},\bar{\boldsymbol{\eta}}_{k}) \leq  \mathbbm{P}_nl(\bar{\beta}_{0k}+\vect{Z}_k\bar{\boldsymbol{\zeta}}_k+\vect{Z}_C\bar{\boldsymbol{\tau}}_{k},\bar{\boldsymbol{\eta}}_{k}) \}$, where $\boldsymbol{\widehat{\zeta}}_k = [\widehat{\beta}_{k},\boldsymbol{\widehat{\beta}}_{\mathcal{I}k}^T]^T$.

\item[\textit{A.10}] 
For each $k$, let $\textbf{S}_k = \mathbb{P}_n\left[[1,\vect{Z}_k,\vect{Z}_C]^T\vect{V}^{-1}[1,\vect{Z}_k,\vect{Z}_C]\right]$. Then, there exists some positive constants $c_7$ and $c_8$ (independent of $k$) and some $0< \kappa < 1/2$, such that
\begin{align*}
    \text{Pr}(\lambda_{\text{min}}(\boldsymbol{S}_k) > c_7) = 1-\mathcal{O}\{\exp(-c_8n^{1-\kappa} )\}.
\end{align*}

\end{itemize}

Conditions A.1-A.8 (except A.2) are equivalent to Condition 2 in \cite{conditional-screening} for multivariate response variables, which hold for most common practical cases. Condition A.2 ensures that the sum of the variances of the covariates is finite, whereas in \cite{conditional-screening} the assumption was made that the marginal variances were equal to one. Conditions A.9 and A.10 are related to the data-generating process. Condition A.9 is required for the sure screening property to hold with an unknown variance parameter $\boldsymbol{\eta}$. Condition A.10 ensures that the matrix $\vect{S}_k$ is positive definite with exponentially high probability, by letting its minimum eigenvalue be bounded away from zero with probability converging to one at an exponential rate.

\section{Appendix}
\label{APP:Proofs}

\subsection{Proof of Proposition 1}
\label{section:A.1}
From \eqref{definiton1}, we have that $E_L(\vect{y}|\vect{Z}_C) = \bar{\beta}_0^{M}\vect{1}+\vect{Z}_C\boldsymbol{\bar{\tau}}^M$ and $E_L(\vect{y}|\vect{Z}_k,\vect{Z}_C) = \bar{\beta}_{0k}\vect{1}+\vect{Z}_k\boldsymbol{\bar{\zeta}}_k+\vect{Z}_C\boldsymbol{\bar{\tau}}_{k}$, where $\boldsymbol{\bar{\zeta}}_k = [\bar{\beta}_k,\boldsymbol{\bar{\beta}}_{\mathcal{I}k}^T]^T$.

$\Leftarrow$:\\
If $E_L(\vect{y}|\vect{Z}_C)=E_L(\vect{y}|\vect{Z}_k,\vect{Z}_C)$ then $\bar{\beta}_0^{M}\vect{1}+\vect{Z}_C\boldsymbol{\bar{\tau}}^M=\bar{\beta}_{0k}\vect{1}+\vect{Z}_k\bar{\boldsymbol{\zeta}}_k+\vect{Z}_C\bar{\boldsymbol{\tau}}_{k}$ which implies $\vect{Z}_k\bar{\boldsymbol{\zeta}}_k = 0$. By A.3, this is true if and  only if $\bar{\boldsymbol{\zeta}}_k = \vect{0}$. 
Because we assume the solution to \eqref{scoreeqs} is unique, then $\bar{\beta}_{0}^M = \bar{\beta}_{0k}$ and $\boldsymbol{\bar{\tau}}^{M}=\boldsymbol{\bar{\tau}}_k$, which means $L_k^R = 0$. 

$\Rightarrow$:\\ 
If  $L_k^R = 0$, then from Condition A.8, $\left\|\left[\bar{\beta}_{0k}-\bar{\beta}_{0}^M,\boldsymbol{\bar{\zeta}}_k^T,(\boldsymbol{\bar{\tau}}_k-\boldsymbol{\bar{\tau}}^M)^T\right]\right\|^2 = 0$, 
which means that $\boldsymbol{\bar{\zeta}}_k= \vect{0}$. By the definition of the conditional linear expectation, $(\bar{\beta}_{0}^M,\vect{0},\boldsymbol{\bar{\tau}}^M)$ is a solution to 
\begin{align}
     \E\Big[\begin{bmatrix}1 & \vect{Z}_C   \end{bmatrix}^T\vect{V}^{-1}(\bar{\beta}_{0}^M\vect{1}+\vect{Z}_C\boldsymbol{\bar{\tau}}^M)\Big] =  \E\Big[\begin{bmatrix}1 & \vect{Z}_C  \end{bmatrix}^T\vect{V}^{-1}\vect{y}\Big].
     \label{score5b}
\end{align}
Similarly, $(\bar{\beta}_{0k},\vect{0},\boldsymbol{\bar{\tau}}_k)$ is a solution to \eqref{scoreeqs}. 
But, note that \eqref{scoreeqs} is equivalent to
\begin{align*}
      \E\left[\begin{bmatrix}1 & \vect{Z}_C \end{bmatrix}^T\vect{V}^{-1}(\bar{\beta}_{0k}\vect{1}+\vect{Z}_k\boldsymbol{\bar{\zeta}}_{k}+\vect{Z}_C\boldsymbol{\bar{\tau}}_{k})\right] &=  \E\left[\begin{bmatrix}1 & \vect{Z}_C \end{bmatrix}^T\vect{V}^{-1}\vect{y}\right], \text{ and } \\
     \E\Big[\vect{Z}_k^T \vect{V}^{-1}(\bar{\beta}_{0k}\vect{1}+\vect{Z}_k\boldsymbol{\bar{\zeta}}_{k}+\vect{Z}_C\boldsymbol{\bar{\tau}}_{k}) \Big]&=  \E\Big[ \vect{Z}_k^T \vect{V}^{-1}\vect{y}\Big]
\end{align*}
Since $\boldsymbol{\bar{\zeta}}_k = \vect{0}$, 
\begin{align}
      \E\left[\begin{bmatrix}1 & \vect{Z}_C  \end{bmatrix}^T\vect{V}^{-1}(\bar{\beta}_{0k}\vect{1}+\vect{Z}_k\boldsymbol{\bar{\zeta}}_{k}+ \vect{Z}_C\boldsymbol{\bar{\tau}}_{k}) \right] \nonumber \\=  \E\left[\begin{bmatrix}1 & \vect{Z}_C  \end{bmatrix}^T\vect{V}^{-1}(\bar{\beta}_{0k}\vect{1}+ \vect{Z}_C\boldsymbol{\bar{\tau}}_{k})\right] \nonumber \\=  \E\left[\begin{bmatrix}1 & \vect{Z}_C \end{bmatrix}\vect{V}^{-1}\vect{y}\right],
     \label{score6b}
\end{align}
implying that a solution to \eqref{score5b} is also a solution to \eqref{scoreeqs}. Since we assume that this solution is unique, we must have that $\bar{\beta}_{0}^M= \bar{\beta}_{0k}, \boldsymbol{\bar{\tau}}^M = \boldsymbol{\bar{\tau}}_k$, i.e. $E_L(\vect{y}|\vect{Z}_C)=E_L(\vect{y}|\vect{Z}_k,\vect{Z}_C)$. \hfill$\square$

\subsection{Proof of Theorem 2}
Let $\boldsymbol{\Omega}_k = \E\left[[1,\vect{Z}_C,\vect{Z}_k]^T\vect{V}^{-1}[1,\vect{Z}_C,\vect{Z}_k]\right]$. We can then partition $\boldsymbol{\Omega}_k$ as follows:
\begin{align}
    \boldsymbol{\Omega}_k = \begin{pmatrix}
    \E\left[[1,\vect{Z}_C]^T\vect{V}^{-1}[1,\vect{Z}_C]\right] & \E\left[[1,\vect{Z}_C]^T\vect{V}^{-1}\vect{Z}_k]\right] \\
\E\left[\vect{Z}_k^T\vect{V}^{-1}[1,\vect{Z}_C]\right] & \E[\vect{Z}_k^T\vect{V}^{-1}\vect{Z}_k ]= 
    \end{pmatrix} = \begin{pmatrix} 
    \boldsymbol{\Omega}_{11} & \boldsymbol{\Omega}_{12} \\ 
    \boldsymbol{\Omega}_{21} &\boldsymbol{\Omega}_{22}\end{pmatrix}.
    \label{eq:definitionOmega}
\end{align}
From the score equations \eqref{scoreeqs} and \eqref{score5b}, \begin{align*}
    \E\left[\begin{bmatrix}1 & \vect{Z}_C \end{bmatrix}^T\vect{V}^{-1}(\bar{\beta}_{0k}\vect{1}+ \vect{Z}_C\boldsymbol{\bar{\tau}}_{k}+\vect{Z}_k\boldsymbol{\bar{\zeta}}_{k})\right] =\E\left[\begin{bmatrix}1 & \vect{Z}_C \end{bmatrix}^T\vect{V}^{-1}(\bar{\beta}_0^M\vect{1}+\boldsymbol{\bar{\tau}}^M\vect{Z}_C)\right]\\ = \E\left[\begin{bmatrix}1 & \vect{Z}_C \end{bmatrix}^T\vect{V}^{-1}\vect{y}\right]
\end{align*}

Let $\boldsymbol{\check{\tau}}_{k} = \boldsymbol{\bar{\tau}}_{k}-\boldsymbol{\bar{\tau}}^M$ and
$\check{\beta}_{0k} = \bar{\beta}_{0k}-\bar{\beta}_0^M$. Then 
\begin{align*}
    \E\left[\begin{bmatrix}1 & \vect{Z}_C \end{bmatrix}^T\vect{V}^{-1}(\check{\beta}_{0k}\vect{1}+ \vect{Z}_C\boldsymbol{\check{\tau}}_{k}+\vect{Z}_k\boldsymbol{\bar{\zeta}}_{k})\right] =0
\end{align*}
Solving this for $(\check{\beta}_{0k},\boldsymbol{\check{\tau}}_{k})$ gives
\begin{align*}
[\check{\beta}_{0k},\boldsymbol{\check{\tau}}_{k}^T]^T = -\boldsymbol{\Omega}_{11}^{-1}\boldsymbol{\Omega}_{12}\boldsymbol{\bar{\zeta}}_k.
\end{align*}
Now, we have that
\begin{align}
    \E[\vect{Z}_k^T\vect{V}^{-1}\{E_L(\vect{y}|\vect{Z}_k,\vect{Z}_C)-E_L(\vect{y}|\vect{Z}_C)\}] =  \E[\vect{Z}_k^T\vect{V}^{-1}\{\check{\beta}_{0k}\vect{1}+\vect{Z}_C\boldsymbol{\check{\tau}}_{k}+\vect{Z}_k\boldsymbol{\bar{\zeta}}_{k}\}] \\
    = \{\boldsymbol{\Omega}_{22}-\boldsymbol{\Omega}_{21}\boldsymbol{\Omega}_{11}^{-1}\boldsymbol{\Omega}_{12}\}\boldsymbol{\bar{\zeta}}_{k}.
    \label{omegasolveforbeta}
\end{align}
We may note that $\boldsymbol{\Omega}_{22}-\boldsymbol{\Omega}_{21}\boldsymbol{\Omega}_{11}^{-1}\boldsymbol{\Omega}_{12}$ is the Schur complement of $\boldsymbol{\Omega}_{22}$ in $\boldsymbol{\Omega}_k$ and is positive semi-definite, because $\boldsymbol{\Omega}_{22}$ is positive definite. From the first part of Condition \eqref{EQ:Cond1}, $\| \boldsymbol{\Omega}_{22}\|\leq M$ for a positive constant $M$. 
Hence,
$$
0 \leq  \|\boldsymbol{\Omega}_{22}-\boldsymbol{\Omega}_{21}(\boldsymbol{\Omega}_{11})^{-1}\boldsymbol{\Omega}_{12}\| \leq \| \boldsymbol{\Omega}_{22} \|\leq M.
$$
Finally, by the second part of Condition (\ref{EQ:Cond1}), we have
\begin{align}
    \|\boldsymbol{\bar{\zeta}}_k\| 
    \geq M^{-1}\left\|\E[\vect{Z}_k^T\vect{V}^{-1}\{E_L(\vect{y}|\vect{Z}_k,\vect{Z}_C)-E_L(\vect{y}|\vect{Z}_C)\}]\right\|
    \geq  c_2m^{1/2}n^{-\kappa},
    \label{EQ:boldzetabound}
\end{align}
where $c_2 = c_0/M$ for $k\in \mathcal{B}$. From Condition $A.8$, there then exists a constant $c_1$ such that 
\begin{align*}
    \lvert L_k^R\rvert \geq c_1mn^{-2\kappa}
\end{align*}
for $k \in \mathcal{B}$.
This completes the proof.
\hfill$\square$

\subsection{Proof of Theorem 3}

In order to establish Theorem 3, we need to first consider some lemmas. 
The first lemma shows a uniform tail bound for the MLEs, whereas the second one gives the tail probability bound for the response vector;
both requires appropriate conditions on the distributions of the covariates as specified in the respective lemmas.

\textbf{Lemma 1:}\textit{
For any $k=1, \ldots, p$, if Conditions A.1-A.9 hold true, then for any $t >0$ and large enough $n$, we have that
\begin{align}
	\text{Pr}\left(\| \widehat{\boldsymbol{\zeta}}_k-\bar{\boldsymbol{\zeta}}_k \|\geq c_5m^{1/2}n^{-\kappa}\right)  
	\leq \exp(-c_6 n^{1-2\kappa}/(k_nK_n)^2) +r_2\exp(-r_3n)+ n\text{Pr}(\Lambda_n^c),
	\label{thm1FanSong}
\end{align}
for some positive constants $c_5$ and $c_6$.
}

\textbf{Proof:} Fix any $k=1, \ldots, p$. 
Note that, under our general uniform formulations, 
$(\vect{y}_{i}, \vect{Z}_{ik}, \vect{Z}_{-ik})_{i=1, \ldots, n}$ are the i.i.d. realizations of the random variables $(\vect{y}, \vect{Z}_{k}, \vect{Z}_C)$;
put $\mathbb{Z}_k=[\vect{1},\vect{Z}_k,\vect{Z}_C] \in \mathbb{R}^{m\times p_m}$, where $p_m = 2m$ and $\boldsymbol{\mu}=[{\beta}_{0k},\boldsymbol{{\zeta}}_{k}^T,\boldsymbol{\tau}_{k}^T]^T$. 
Then, Assumptions A.1--A.6 in Theorem 1 in \cite{bratsberg2023exponential} hold for each marginal model where the loss function is the log-likelihood with only covariates $\mathbb{Z}_k$ included. 
Therefore, the result follows for regression coefficients $\boldsymbol{\mu}$ by an application of Theorem 1 of \cite{bratsberg2023exponential}  with $1+t = \tilde{K}n^{1/2-\kappa}/16C^{1/2}k_n$. Thus, the results follow from the fact that $\text{Pr}\left(\| \widehat{\boldsymbol{\zeta}}_k-\bar{\boldsymbol{\zeta}}_k \|\geq u \right) \leq \text{Pr}\left(\| \widehat{\boldsymbol{\mu}}-\bar{\boldsymbol{\mu}} \|\geq u\right)$ for any positive $u$, where $\widehat{\boldsymbol{\mu}} =[\widehat{\beta}_{0k},\widehat{\boldsymbol{\zeta}}_{k}^T,\widehat{\boldsymbol{\tau}}_{k}^T]^T$ and $\bar{\boldsymbol{\mu}} =[\bar{\beta}_{0k},\bar{\boldsymbol{\zeta}}_{k}^T,\bar{\boldsymbol{\tau}}_{k}^T]^T.$
\hfill$\square$ 

\bigskip
\textbf{Lemma 2:}
\textit{  If Condition A.4 holds, then we have 
 \begin{align*}
    \text{Pr}(\|\vect{y}\|_\infty \geq m_0t^\alpha/s_0 ) \leq  s_1\exp(-m_0t^\alpha), ~~~~~\text{ for any } t>0.
\end{align*}
}
\textbf{Proof:}
Let $\vect{y} = [y_1, \ldots, y_m]^T$. 
By the exponential Chebyshev's inequality, as in the proof of Lemma 1 of \cite{sis_in_generalized_linear_models}, we can show that
\begin{eqnarray}
\text{Pr}(|y_j| \geq u) &\leq& \exp(-s_0u)\E[\exp(s_0y_j)+\exp(-s_0y_j)]\\&=& \exp(-s_0u)\E\{\E[\exp(s_0y_j)\lvert \vect{x}_j^T\boldsymbol{\beta}^*,\vect{q}_j^T\vect{b}]+\E[\exp(-s_0y_j)\lvert \vect{x}_j^T\boldsymbol{\beta}^*,\vect{q}_j^T\vect{b}]\}
\nonumber\\
&\leq& \exp(-s_0u)\Bigg[\E\Big\{\exp\big[(\vect{x}_j^T\boldsymbol{\beta}^*+\vect{q}_j^T\vect{b}+s_0)^2/2-(\vect{x}_j^T\boldsymbol{\beta}^*+\vect{q}_j^T\vect{b})^2/2\big]\Big\}
\nonumber\\
&&~~~~~~~~+ \E\Big\{\exp\big[(\vect{x}_j^T\boldsymbol{\beta}^*+\vect{q}_j^T\vect{b}-s_0)^2/2-(\vect{x}_j^T\boldsymbol{\beta}^*+\vect{q}_j^T\vect{b})^2/2\big]\Big\}\Bigg],
~~~\text{ for } j=1, \ldots, m.
\nonumber
\end{eqnarray}
Therefore, we get
\begin{align*}
\text{Pr}(\|\vect{y}\|_\infty \geq u) &\leq \sum_{j=1}^m\text{Pr}(|y_j| \geq u)\\
&\leq  \exp(-s_0u)\sum_{j=1}^m\Bigg[\E\Big\{\exp\big[(\vect{x}_j^T\boldsymbol{\beta}^*+\vect{q}_j^T\vect{b}+s_0)^2/2-(\vect{x}_j^T\boldsymbol{\beta}^*+\vect{q}_j^T\vect{b})^2/2\big]\Big\}
	\\
	&~~~~~~~~~~~~~~~+ \E\Big\{\exp\big[(\vect{x}_j^T\boldsymbol{\beta}^*+\vect{q}_j^T\vect{b}-s_0)^2/2-(\vect{x}_j^T\boldsymbol{\beta}^*+\vect{q}_j^T\vect{b})^2/2\big]\Big\}\Bigg] \\
&\leq s_1\exp(-s_0u),
\end{align*}
where the last step follows from Condition A.4. Then, the desired result is obtained by letting $u = m_0t^\alpha/s_0$.
\hfill$\square$

\bigskip
\subsubsection*{Proof of Theorem 3:}

The idea is to bound $\widehat{L}_k^R$ from below to show the strength of the signals.  By Taylor expansion, we have that
\begin{align*}
    2\widehat{L}_k^R = \left[\widehat{\beta}_{0}^M-\widehat{\beta}_{0k},\boldsymbol{\widehat{\zeta}}_k^T,(\boldsymbol{\widehat{{\tau}}}^M-\boldsymbol{\widehat{{\tau}}}_k)^T\right]\vect{L}''_k\left[\widehat{\beta}_{0}^M-\widehat{\beta}_{0k},\boldsymbol{\widehat{\zeta}}_k^T,(\boldsymbol{\widehat{{\tau}}}^M-\boldsymbol{\widehat{{\tau}}}_k)^T\right]^T \\\geq \lambda_{k,min}\|\widehat{\boldsymbol{\zeta}}_k\|^2
\end{align*}
where $\lambda_{k,min}$ is the minimum eigenvalue of
\begin{align*}
    \vect{L}''_k :=  \mathbb{P}_n\left[\begin{bmatrix}1 & \vect{Z}_k & \vect{Z}_C\end{bmatrix}^T\vect{V}^{-1} \begin{bmatrix}1 & \vect{Z}_k& \vect{Z}_C\end{bmatrix}\right],
\end{align*} 
which by Condition A.10 is bounded according to
\begin{align}
 \text{Pr}(\lambda_{k,min} > c_7) = 1-\mathcal{O}\{\exp(-c_8n^{1-\kappa} )\}.
 \label{EQ:minlambdabound}
\end{align}

Now, we bound $\|\widehat{\boldsymbol{\zeta}}_k\|$. By \eqref{EQ:boldzetabound},
\begin{align*}
	\min_{k \in \mathcal{B}}\| \boldsymbol{\bar{\zeta}}_k \|\geq c_2m^{1/2}n^{-\kappa}.
\end{align*}

By applying Lemma 1 over all $k \in \mathcal{B}$ and the union bound of probability, we have
\begin{align}
    \text{Pr}\Big(\max_{k \in \mathcal{B}} \|\widehat{\boldsymbol{\zeta}}_k-\bar{\boldsymbol{\zeta}}_k\|\leq c_2m^{1/2}n^{-\kappa}/2\Big) 
	\geq 1- 
	s\exp(c_4n^{1-2\kappa}(k_nK_n)^{-2})-sn\text{Pr}(\Lambda_n^c),
	\label{EQ:lemma1bound}
\end{align}
for some constant $c_4$.
Further, by Lemma 2, we have 
\begin{align*}
    \text{Pr}(\|\vect{y} \|_\infty > t ) \leq  s_1\exp(-s_0t).
\end{align*}
Thus, using this tail probability bound and Condition A.5, we get
\begin{align}
    \text{Pr}(\Lambda_n^c) &\leq \text{Pr}(\|[\vect{1},\vect{Z}_k,\vect{Z}_C]\|_1 > K_n)+\text{Pr}(\| \vect{y} \|_\infty > r_0K_n^{\alpha}/s_0) 
     \leq r_1\exp(-r_0K_n^{\alpha}),
    \label{EQ:Lambdabound}
\end{align}
where $K_n \to \infty$ as $n \to \infty$.
Hence, 
\begin{align*}
    \text{Pr}\Big(\min_{k \in \mathcal{B}} \|\widehat{\boldsymbol{\zeta}}_k\|\geq c_2m^{1/2}n^{-\kappa}/2\Big) = 1-\mathcal{O}\big(s\exp(c_4n^{1-2\kappa}(k_nK_n)^{-2})+snr_1\exp(-r_0K_n^{\alpha})\big).
\end{align*}

Combining the bound above with \eqref{EQ:minlambdabound} gives
\begin{align*}
    \text{Pr}\Big(\min_{k \in \mathcal{B}}  \widehat{L}_k^R \geq c_7c_2^2mn^{-2\kappa}/8\Big)= 1-\mathcal{O}\big(s\exp(c_4n^{1-2\kappa}(k_nK_n)^{-2})+snr_1\exp(-r_0K_n^{\alpha})\big).
\end{align*}
Hence, by setting $\nu_n = c_3mn^{-2\kappa}$ for $c_3 <c_7c_2^2/8 $ we have that $\mathcal{B} \subset \widehat{\mathcal{B}}$ with the probability tending to one exponentially fast. This completes the proof.
\hfill $\square$

\end{document}